\documentclass[acmsmall,printacmref=false,printfolios=false,nonacm=true]{acmart}

\newtheorem{mydef}{Definition}

\usepackage{multirow}
\usepackage{yfonts}
\usepackage{longtable}
\usepackage{arydshln}
\usepackage{makecell}
\usepackage{algorithm}
\usepackage{algorithmic}
\usepackage{enumitem}
\usepackage{amsmath}
\usepackage{adjustbox}

\usepackage{subcaption}

\AtBeginDocument{%
  \providecommand\BibTeX{{%
    \normalfont B\kern-0.5em{\scshape i\kern-0.25em b}\kern-0.8em\TeX}}}

\copyrightyear{2022}
\acmYear{2022}
\acmDOI{10.1145/1122445.1122456}

\begin{document}

\title{Deconfounded Causal Collaborative Filtering}

\author{Shuyuan Xu}
 \affiliation{
  \institution{Rutgers University}
  \country{New Brunswick, NJ, US}
    }
 \email{shuyuan.xu@rutgers.edu}

\author{Juntao Tan}
 \affiliation{
  \institution{Rutgers University}
  \country{New Brunswick, NJ, US}
    }
 \email{juntao.tan@rutgers.edu}

\author{Shelby Heinecke}
 \affiliation{
  \institution{Salesforce Research}
  \country{Palo Alto, CA, US}
    }
 \email{shelby.heinecke@salesforce.com}

\author{Vena Jia Li}
 \affiliation{
  \institution{Meta}
  \country{Menlo Park, CA, US}
    }
 \email{vena900620@gmail.com}

\author{Yongfeng Zhang}
 \affiliation{
  \institution{Rutgers University}
  \country{New Brunswick, NJ, US}
    }
 \email{yongfeng.zhang@rutgers.edu}
 
\begin{CCSXML}
<ccs2012>
   <concept>
       <concept_id>10010147.10010257</concept_id>
       <concept_desc>Computing methodologies~Machine learning</concept_desc>
       <concept_significance>500</concept_significance>
       </concept>
   <concept>
       <concept_id>10002951.10003317.10003347.10003350</concept_id>
       <concept_desc>Information systems~Recommender systems</concept_desc>
       <concept_significance>500</concept_significance>
       </concept>
 </ccs2012>
\end{CCSXML}

\ccsdesc[500]{Computing methodologies~Machine learning}
\ccsdesc[500]{Information systems~Recommender systems}

\begin{abstract}

Recommender systems may be confounded by various types of confounding factors (also called confounders) that may lead to inaccurate recommendations and sacrificed recommendation performance. Current approaches to solving the problem usually design each specific model for each specific confounder. However, real-world systems may include a huge number of confounders and thus designing each specific model for each specific confounder could be unrealistic. More importantly, except for those ``explicit confounders'' that experts can manually identify and process such as item's position in the ranking list, there are also many ``latent confounders'' that are beyond the imagination of experts. For example, users' rating on a song may depend on their current mood or the current weather, and users' preference on ice creams may depend on the air temperature. Such latent confounders may be unobservable in the recorded training data. To solve the problem, we propose Deconfounded Causal Collaborative Filtering (DCCF). We first frame user behaviors with unobserved confounders into a causal graph, and then we design a front-door adjustment model carefully fused with machine learning to deconfound the influence of unobserved confounders. Experiments on real-world datasets show that our method is able to deconfound unobserved confounders to achieve better recommendation performance.

\end{abstract}

\keywords{Recommender Systems; Deconfounded Recommendation; Causal Analysis}

\maketitle
% \pagestyle{plain}
% \pagestyle{empty}
% \thispagestyle{empty}

% \vspace{-1ex}
\section{Introduction}\label{sec:intro}
\begin{figure}[t]
    \centering
    \includegraphics[width=0.9\linewidth]{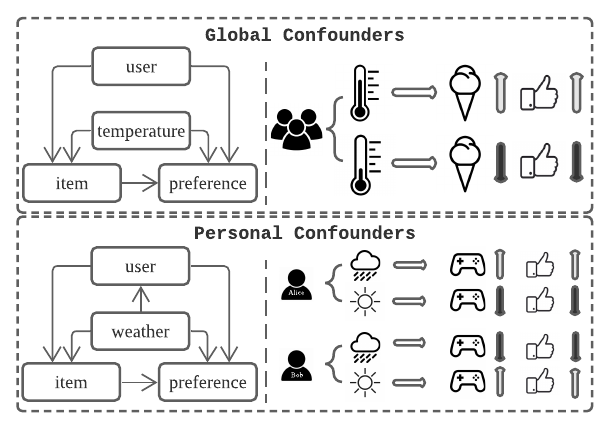}
    \vspace{-15pt}
    \caption{Examples of global and personal confounders.
    % The upper dashed box shows an example of global confounders. We consider the temperature as a global confounder in a causal graph. More specifically, Most users tend to purchase ice cream and like it at high temperature and vice versa at low temperature. The lower dashed box shows an example of personal confounders. We consider the weather as a personal confounder. Concretely, different weather will affect the user's behavior and preference, but the specific impact varies from person to person.
    }
    \label{fig:confounderexample}
    \vspace{-17pt}
\end{figure}

Recommender systems occupy an expanding role in daily decision making, such as e-commerce, video streaming and social media. Most of the existing recommendation models learn from the collective historical interactions to achieve good recommendation performance. However, the collected data may have been affected by various types of confounders, i.e., variables that affect both treatment assignment (which item to be exposed to the user) and outcome (whether the user likes the item) \cite{pearl2016causal,wang2020causal,li2022causal}, which may lead to spurious correlation \cite{xu2023survey,wang2021deconfounding}. As a result, the collected data may not represent the accurate preference of users. Therefore, 
% For example, the system tends to recommend popular items and users tend to consume popular items. 
recommendation algorithms that are trained with confounded data may result in inaccurate recommendations and sacrificed performance.

The existence of confounders naturally leads to a question: can we develop models to mitigate the influence of confounders so as to estimate more accurate user-item preferences? 
% Current approaches to answering this question usually identify a specific confounder based on researchers' expertise and design a specific model to tackle this confounder. 
To answer this question, current approaches usually identify one specific confounder based on researchers' expertise. Then, they design specific models to tackle the pre-identified confounding problem. For instance, % the popularity confounder will result in biased recommendation performance as shown in the previous example. 
many models are proposed to handle the popularity confounders \cite{krishnan2018adversarial,zhang2021causal,zhu2021popularity,zhu2021popularity2,wei2021model}, click confounders \cite{wang2021deconfounding,schnabel2016recommendations,wang2016learning}, exposure confounders \cite{abdollahpouri2020multi,gupta2021correcting,saito2020unbiased,yang2018unbiased}, etc.
% Even if more confounders are identified, the inference is valid if we have accounted for all confounders \cite{rosenbaum1983central}. 
Existing efforts have performed well on combating these manually identified confounders. However, real-world systems may involve a huge number of confounders, which makes it unrealistic to design each specific model for each specific confounder.
% or design a model accounting all identified confounders. 
Moreover, the confounders are not limited to the ``explicit confounders'' that can be manually identified by experts based on their knowledge, but also include many ``latent confounders'' that can be hardly identified or measured. For example, a user's interaction and rating with music recommendations can be affected by his/her current mood or spirit, however, the mood or spirit cannot be quantitatively measured or logged in the training data. The large number of confounders and the existence of latent confounders make it important to develop ``universal'' deconfounded recommendation models that are agnostic to certain types or certain numbers of confounders. It is worth mentioning that deconfounding is similar with debiasing in some cases but they are different concepts. Confounding requires the confounders to affect both treatments (exposures) and outcomes (feedbacks), while biases such as exposure bias or selection bias only result in non-uniform exposures \cite{xu2023survey}. More specifically, if a variable affects both exposures and feedbacks, it creates the discrepancy between the observed and true feedbacks, which is the confounding issue. Otherwise, if a variable only leads to non-uniform exposures, it is the normal biasing issues.
% Indeed, most biases can be understood with mild cause assumptions and additional confounding factors in the causal graph \cite{chen2020bias}. However, when we involve unobserved confounders, some issues cannot be explained by biases. For example, the rating provided by a user can be affected by his/her mood at the time, which may not accurately reflect the user's true preference. 

% To tackle this problem, we first examine the nature of latent confounders. 
Broadly speaking, the confounders can be classified into global confounders and personal confounders, as shown in Figure \ref{fig:confounderexample}. Global confounders are those that influence preferences of all users in a consistent way, while personal confounders are those that affect different users differently. For example, the air temperature can be a global confounder for users' preferences on ice creams---during normal times, some users tend to like ice creams while others tend to dislike ice creams, however, an extremely high temperature may increase both groups' preference on ice creams. 
% On the other hand, the weather can be a personal confounder for users' preferences on household PlayStations such as Xbox. For example, some users like sunny days and they choose to take outdoor activities, as a result, their preferences on PlayStations decrease when sunny while increase when rainy since they have to stay at home. However, some other users may dislike sunny days due to skin protection reasons and thus they choose to stay at home when sunny with an increased preference on PlayStations, while their preference on PlayStations decreases when rainy since they choose to go out when the UV exposure is low. 
On the other hand, the weather can be a personal confounder for users' preferences on game consoles, which means it may have different effects for different users. For example, some users preferences on game consoles might decrease in sunny days while increasing in rainy days. This is because they choose to take outdoor activities in sunny days but stay at home in rainy days. However, for other users, their preference on game consoles may increase in sunny days because they have certain skin disease and avoid going outside when it's sunny.
This complex nature of confounders makes it challenging to design deconfounded recommendation models.

Fortunately, causal inference techniques---especially front-door adjusted models---make it possible to design a unified deconfounding model for both global and personal unobserved confounders \cite{xu2023survey,pearl2016causal}. 
% For both ``explicit confounders'' and ``latent confounders'', confounders can be categorized into global confounders and personal confounders. We show some examples in Figure \ref{fig:confounderexample}. More generally, the effect of global confounders is consistent across most users while the effect of personal confounders varies for different users.
% Recommender Systems (RS) aim to recommend items that users will like. Current methods usually consider it as a prediction task, concretely, they learn users' preferences based on observational data to make predictions on items that users have not interacted with and make recommendation based on those predictions. Those methods have achieved great successes. However, the spirit of recommender systems is not only to estimate user-item associative relationships, but to estimate the effect if we recommend something to users. 
Actually, the essence of recommendation is trying to answer a ``what if'' question: what would happen if we recommend a certain item to a target user? Inspired by causal collaborative filtering \cite{xu2023causal,xu2022dynamic,xu2022causal}, this ``what if'' question can be represented as $P(y|u,do(v))$, where $u,v$ is a user-item pair and $y$ is the estimated preference score.
In particular, the user, item, and preference can be formulated into a causal graph, while unobserved global or personal confounders present possible connections to the user, item and preference nodes (Figure \ref{fig:confounderexample}).
% both global and personal confounders can be represented as variables 
% Combining global confounders and personal confounders, we first design a causal graph to describe the recommendation scenario with unobserved confounders. 
Technically, we identify inherent item features as a mediator variable $M$ between item and preference that is independent from the influence of confounders (Figure \ref{fig:causal}, more details later). Based on mediator analysis, we are able to estimate the deconfounded user-item preference $P(y|u,do(v))$ based on front-door adjustments rooted in causal theory \cite{pearl2016causal}, where $u,v$ is a pair of user and item, $y$ is the preference score between them and the $do$-operation is used to model the interventions. Intuitively, the deconfounded preference is used to represent the causal preference if we intervene to recommend item $v$ instead of passively observing item $v$ in training data. Considering the large scale of items in recommender systems, it is impractical to strictly apply the front-door adjustment into the model, therefore, we design a sample-based approach to make it calculable (more details will be shown in Section \ref{sec:model}).  
% Based on our designed causal graph, we then propose a deconfounded recommender system from a causal view. More specifically, the proposed deconfounded model aims to estimate $P(y|u,do(v))$ where $u,v$ is a user-item pair and $y$ is the estimated preference score. Technically, to estimate $P(y|u,do(v))$ with the existence of unobserved confounders, we apply the front-door adjustment, which is a fundamental causal technique to solve confounding problems. 
% To improve the efficiency of estimation under front-door adjustments, we propose a sample based front-door adjustment method for recommendation. 
% \yz{Todo: add some technical challenge}
In the experiments, we compare our model with both state-of-the-art deconfounded recommendation algorithms and traditional association-based  recommendation models. The results show that our deconfounded algorithm achieves better recommendation performance than all of the baselines. In summary, we list our key contributions as follows:

\begin{itemize}[leftmargin=*]
    \item We design a causal graph to describe the data generation process and represent the effects of unobserved confounders in the recommendation scenario.
    % We design a causal graph depicting the recommendation scenario with unobserved confounders, including global confounders and personal confounders.
    \item To mitigate the effects of unobserved confounders which are not directly measurable, we adapt the front-door adjustment into the proposed deconfounded causal recommnedation model. 
    % To the best of our knowledge, this is the first time front-door adjustment is leveraged for causal recommendation.
    % We propose a deconfounded recommender system applying the front-door adjustment to deconfound the effect of unobserved confounders, and design a sample-based approach to improve efficiency.
    \item We design a sample-based approach integrated with exposure models to make the front-door adjustment calculable despite the large item space.
    % to further improve the efficiency of our recommendation model. \yz{Todo: replace with another technical contribution}
    % Comprehensive ablation studies are conducted to explore the relationship between the recommendation performance and the different sample strategies.
    \item Experiments on three real-world datasets show that our deconfounded model outperforms existing deconfounded recommendation models and traditional association-based models.
\end{itemize}

The remainder of this paper is organized as follows. We discuss the related works in Section \ref{sec:related}. In Section \ref{sec:preliminary}, we introduce some notations, basic concepts and theorems to help readers gain a better understanding of the fundamentals. We introduce our proposed model in Section \ref{sec:model}. In Section \ref{sec:experiments}, we experiment on real-world datasets and make discussions. Finally, we conclude the work and discuss future directions in Section \ref{sec:conclusion}.

% \vspace{-1ex}
\section{Related Work}\label{sec:related}
% We review the related works on deconfounded recommender system and causal inference in recommendations.
% \subsection{Deconfounded recommendation}
% \cite{wang2020causal}

% Although there are some related works on confounding problem in other communities, in this section, we only focus on the recommender systems (RS) community and introduce some related works. 

% Logic of related work: deconfounded recommendation is important -> traditional method: IPS -> IPS has problems -> recent methods: causal graph -> but current causal graph methods only work with one specific confounder -> there are some other works such as AutoDebias, but they rely on unbiased data.

Deconfounded recommender systems aim to remove or reduce the unwanted effect of confounders, which is important to the accurate estimation of user preferences \cite{xu2023survey}.
% Debiasing in RS is a related area that focuses on developing techniques to mitigate the effects of specific biases, represented as specific confounders in some works.
% Compared with deconfounding problem in RS, bias and debias in RS is a similar and overlapped area. Related works put their efforts into handling one specific bias in RS. 
% Therefore, approaches to debiasing can be basically categorized based on addressed bias. 
% The data of user interactions are observational rather than experimental, which makes it easy to introduce bias in the data. 
Among the large number of methods mitigating the confounding effects, causal technique plays an important role. Inverse Propensity Score (IPS) based method is an important approach for deconfounding. The basic idea is to re-weight the observations with inverse propensity scores
% It is capable of mitigating many types of bias, 
so as to mitigate the influence of selection bias \cite{schnabel2016recommendations,wang2016learning}, exposure bias \cite{yang2018unbiased,saito2020unbiased,wang2022unbiased}, position bias \cite{agarwal2019estimating,vardasbi2020cascade,guo2020debiasing,joachims2017unbiased}, etc. IPS-based methods are well used approaches for deconfounded learning with observational data, but they still suffer from some known issues. For example, the performance of IPS-based methods highly depends on the accurate estimation of propensity scores \cite{saito2020asymmetric}
% and the unbiasedness is guaranteed only when the true propensity scores are available . 
and usually suffers from the high variance of the propensity scores \cite{bonner2018causal}. 

% Recent: clickbait paper. leveraging.

Besides IPS, leveraging causal graph is a powerful approach to deconfounded recommendation. Many existing methods construct a specific causal graph based on certain assumptions to incorporate specific confounders into the data generation process, and then apply causal inference techniques to mitigate the confounding bias. 
Just to name a few as examples, Zhang et al. \cite{zhang2021causal} assumed that item popularity affects the item exposure and user interaction, and constructed a causal graph incorporated with item popularity to estimate the user-item preferences;
Qiu et al. \cite{ruihong2021causalrec} observed visual bias in visual-aware recommendation and that users' attention to visual features does not always reflect the real preference, and thus developed a causal graph to remove the effect of visual confounders; Li et al. \cite{li2021towards} noticed that users' sensitive features may lead to unfair recommendations and thus developed causal graphs to deconfound the influence of sensitive features in recommendation; Li et al. \cite{li2022causal} achieved causal feature selection in factorization machines so as to enhance the robustness
of recommendation when the distributions of training data and testing data are different;
Wang et al. \cite{wang2021deconfounding} identified that user clicks may be influenced by item popularity and proposed a deconfounding method to alleviate the amplification of popularity bias.
Many other research works are conducted to address different types of confounders, including but not limited to item popularity \cite{zhu2021popularity,zhu2021popularity2,wei2021model,ge2021towards,abdollahpouri2020connection,krishnan2018adversarial}, item exposure \cite{abdollahpouri2020multi,gupta2021correcting,saito2020unbiased,yang2018unbiased,wang2020causal,wang2019blessings}, user selection \cite{ovaisi2020correcting,marlin2007collaborative,wang2021deconfounding,schnabel2016recommendations,wang2016learning}, and ranking positions \cite{agarwal2019estimating,qin2020attribute,ai2018unbiased,hu2019unbiased,oosterhuis2020policy,wu2021unbiased}. Furthermore, causal and counterfactual reasoning has also been adopted to tackle many other issues in recommender systems such as explainability \cite{tan2021counterfactual,tan2022learning,ji2023counterfactual}, fairness \cite{li2021towards}, robustness \cite{chen2023dark,li2022causal}, adversarial attacks \cite{chen2023dark}, data augmentation \cite{wang2021counterfactual,xiong2021counterfactual,chen2022data}, and evaluating intelligent systems \cite{ge2021counterfactual}.
% \yz{add citation}
% apply causal intervention technique to remove the effect of item popularity. 

% Existing approaches to deconfounded recommender systems mostly design specific deconfounding models for specific confounders. For example, the item popularity may influence the recommendation mechanism, promote already popular items, and create Matthew effects \cite{ge2021towards,abdollahpouri2020connection}, as a result, many models are developed to mitigate the influence of item popularity in recommendation \cite{zheng2021disentangling,krishnan2018adversarial,zhang2021causal}; the position of an item in a recommendation list can also; user clicks 
% For example, in explicit feedback data, selection bias \cite{marlin2007collaborative}, also refer to missing-not-at-random (MNAR) problem, makes observed ratings are not a representative sample of all ratings. 
% In implicit feedback data, unobserved interactions do not always represent negative preference, which leads to exposure bias, also refers to ``selection bias'' \cite{wang2016learning, ovaisi2020correcting} or ``popularity bias'' \cite{zheng2021disentangling} in some works. 
% As we mentioned previously, there can be bias in observational data. 
However, due to the large number of confounders and the existence of unobserved latent confounders, it is unrealistic to design each specific model to tackle each specific confounder. 
% may not always exist and can be difficult to collect.
Some recent works took advantage of a subset of unbiased data as supervision to remove confounding effects \cite{yuan2019improving}. However, unbiased data may not always exist and collecting unbiased data can be difficult since the randomized recommendations involved may hurt the user experience. Some recent deconfounded methods rely on certain user information such as user social network in \cite{gao2021deconfounding}. However, user information may be sensitive and not available sometimes. A recent work by \citeauthor{zhu2022mitigating} considers click behavior as mediator and designs a multi-task learning framework that simultaneously learns the probability of click and purchase to mitigate confounding effects \cite{zhu2022mitigating}. \citet{shang2019environment} designed a confounder policy to handle the hidden confounder for environment reconstruction on reinforcement learning based recommendation. However, this model is not suitable for general settings.

As a result, a broad-spectrum deconfounded recommendation model to mitigate the confounding effects based on observed data is urgently needed.
% which is the key purpose of this paper. 
% As we will show later, causal 
% and it can only is only a very small portion of massive online traffic. 
% In the interests of brevity, we only discussed few types of bias and causal solutions, interested readers may refer to the survey \cite{chen2020bias} for more detail about bias and debias in recommender systems.
% \cite{wei2021model}
% Most debiasing methods mainly target one specific bias, therefore, the solutions vary for different types of bias. A recent approach \cite{Chen2021AutoDebiasLT} is proposed to handle all data biases, however, it requires unbiased data which is hard to access.
% Given most biases can be understood with mild cause assumptions and additional confounding factors in the causal graph \cite{chen2020bias}, mitigating the effect of unobserved confounders will address most biases. 
% Wang et al. \cite{wang2020causal,wang2019blessings} use the exposure data to train an exposure model to substitute the unobserved confounders.
As we will show later, we incorporate the unobserved confounders into the causal graph and leverage front-door adjustments to mitigate the effect of confounders, which is able to deconfound recommender systems based on observed data without the need to enumerate confounders or collect unbiased datasets.

% \yz{the following is copied from introduction.}
% There are previous works that try to address the problem of confounders in recommender systems. Inverse Propensity Score (IPS) (also called inverse propensity weighting) is a well adapted technique for this problem \cite{schnabel2016recommendations,saito2020unbiased}. These methods design a propensity score estimator to approximate the probability of exposure and use the inverse of propensity scores to correct the preference learned from observed feedback. To get more accurate propensity scores, it may require either feedback of a randomized trial or external user item covariates. Additionally, Considering the huge number of items and the high sparsity of the data, the IPS-based methods may not properly handle large shifts in exposure probability. In other words, lower exposure probability may lead to higher prediction scores. Except for the IPS-based models, Wang et al. \cite{wang2020causal} apply the method introduced in \cite{wang2019blessings} to learn a deconfounded recommender system. 
% It first builds an exposure model to estimate the items that each user is likely to consider. This model is trained with the exposure data.
% It then uses this exposure model to estimate a substitute for the unobserved confounders. Finally, it learns a matrix factorization model with the substitute confounders as a recommendation model.

% \vspace{-4pt}
\section{Preliminaries}\label{sec:preliminary}

As we introduced in Section \ref{sec:intro}, our model is able to estimate the deconfounded user-item preference $P(y|u,do(v))$ by leveraging the designed causal graph. We will first introduce some basic concepts in causal theory.

\begin{mydef}
(Causal Graph) \cite[p.35]{pearl2016causal} A causal graph is a directed acyclic graph (DAG) $\mathcal{G}=(U,E)$, which captures the direct causal relationships among the variables.
\end{mydef}

The key component of our deconfounded estimation is $do$-operation, which models the interventions, in the estimated preference $P(y|u,do(v))$. We introduce intervention as follows:

\begin{mydef}
(Intervention) \cite[p.55]{pearl2016causal} We distinguish between cases where a variable $X$ takes a value $x$ naturally and cases where we fix $X=x$ by denoting the later $do(X=x)$. So $P(Y=y|X=x)$ is the probability that $Y=y$ conditioned on finding $X=x$, while $P(Y=y|do(X=x))$ is the probability that $Y=y$ when we intervene to make $X=x$. Similarly, we write $P(Y=y|do(X=x),Z=z)$ to denote the conditional probability of $Y=y$, given $Z=z$, in the distribution created by the intervention $do(X=x)$.
% The distribution of variable $Y$ under intervention $X=x$---noted as $P(Y=y|do(X=x))$ or simply $P(y|do(x))$---is the distribution of $Y$ when we intervene to fix $X=x$ in the causal model, in contrast to $P(Y=y|X=x)$ where . $P(Y=y|do(X=x),Z=z)$ simplified as $P(y|do(x),z)$. Conditional intervention.
\end{mydef}

Given a causal graph, there are some methods to estimate the desired interventional probabilities in causal theory. The back-door adjustment and the front-door adjustment are two fundamental theorems in causal inference that can be used to estimate the deconfounded user-item preference using the observational data. However, according to our designed causal graph (will introduce in Section \ref{sec:graph}), the back-door adjustment is not applicable for our causal graph because the confounders are unobserved. We will first introduce the back-door adjustment and the front-door adjustment in causal inference. 
% Back-door adjustment has been used for many deconfounding methods \yz{TODO}. 
Additionally, for better understanding, we will mathematically illustrate why back-door adjustment is not applicable in our setting and why we use front-door adjustment instead to estimate the deconfounded user-item preference.
% \begin{mydef}
% (Structural Causal Models) \cite[p.26]{pearl2016causal} A structural causal model (SCM) $M$ consists of two set of variables $U$ and $V$, and a set of functions $f$ that assign a value to each variable in $V$ based on other variables in the model. Here $U$ are exogenous variables that no explanatory mechanism is encoded.
% \end{mydef}

% \begin{mydef}
% (Causal Graph) \cite[p.35]{pearl2016causal} A causal graph is a directed acyclic graph (DAG) $\mathcal{G}=(\{U,V\},E)$, which captures the relationships among the variables in the corresponding SCM.
% \end{mydef}

\begin{figure}[t]
    \centering
    \includegraphics[width=0.8\linewidth]{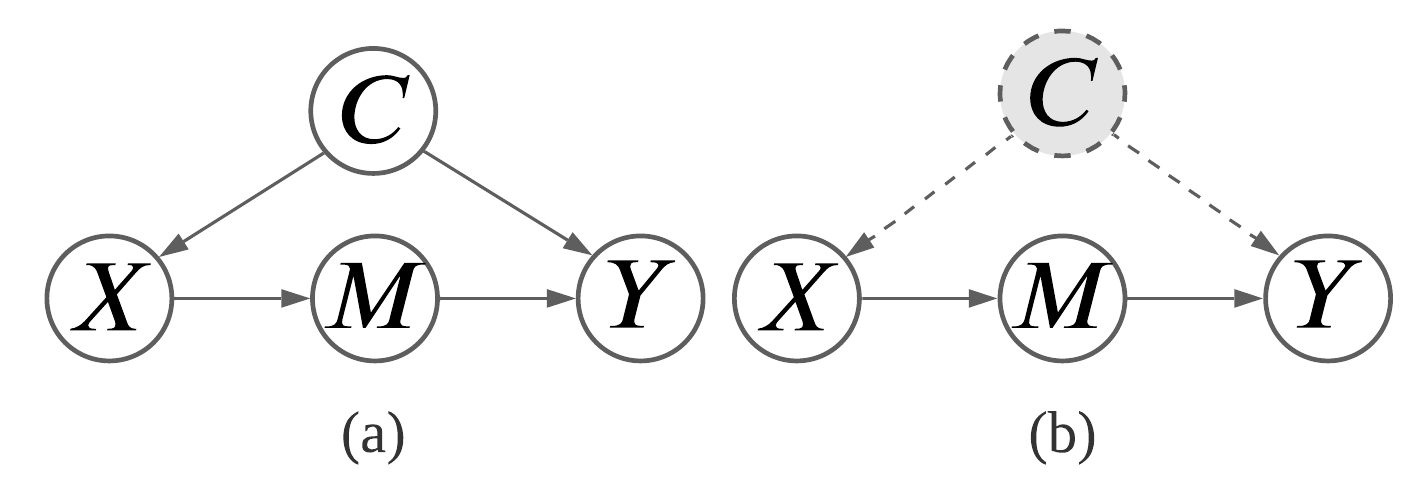}
    \vspace{-10pt}
    \caption{$X$ represents treatment variable, $Y$ represents outcome variable, $M$ represents mediator variable, $C$ represents confounder variable. Figure (a) shows variable $C$ as observed and measurable, while in Figure (b), variable $C$ is unobserved or unmeasurable.}
    \label{fig:adjustment}
    \vspace{-5pt}
\end{figure}

% \begin{mydef}
% (Intervention) \cite[p.55]{pearl2016causal} We distinguish between cases where a variable $X$ takes a value $x$ naturally and cases where we fix $X=x$ by denoting the later $do(X=x)$. So $P(Y=y|X=x)$ is the probability that $Y=y$ conditioned on finding $X=x$, while $P(Y=y|do(X=x))$ is the probability that $Y=y$ when we intervene to make $X=x$. Similarly, we write $P(Y=y|do(X=x),Z=z)$ to denote the conditional probability of $Y=y$, given $Z=z$, in the distribution created by the intervention $do(X=x)$.
% % The distribution of variable $Y$ under intervention $X=x$---noted as $P(Y=y|do(X=x))$ or simply $P(y|do(x))$---is the distribution of $Y$ when we intervene to fix $X=x$ in the causal model, in contrast to $P(Y=y|X=x)$ where . $P(Y=y|do(X=x),Z=z)$ simplified as $P(y|do(x),z)$. Conditional intervention.
% \end{mydef}

\begin{mydef}
(Back-door Criterion) \cite[p.61]{pearl2016causal} Given an ordered pair of variables $(X,Y)$ in a causal graph $\mathcal{G}$, a set of variables $Z$ satisfies the back-door criterion with respect to $(X,Y)$ if $Z$ satisfies the following conditions: 
% (1) No node in $Z$ is a descendant of $X$ and (2) $Z$ blocks every path between $X$ and $Y$ that contains an arrow into $X$.
\begin{itemize}
\renewcommand\labelitemi{--}
    \item No node in $Z$ is a descendant of $X$;
    \item $Z$ blocks every path between $X$ and $Y$ that contains an arrow into $X$.
\end{itemize}
\end{mydef}

With the help of a set of variables that satisfy the back-door criterion, we can adjust for the effect of measured confounders. We take the causal graph in Figure \ref{fig:adjustment}(a) as an example. Considering the treatment variable $X$ and the outcome variable $Y$, we want to estimate the effect of $X$ on $Y$, which is denoted as $P(Y=y|do(X=x))$. Due to the existence of confounder $C$, we cannot conclude that $P(Y=y|do(X=x))=P(Y=y|X=x)$. However, since variable $C$ satisfies the back-door criterion, we use it to adjust the effect, in other words, we are accounting for and measuring all confounders \cite{rosenbaum1983central,pearl2016causal}. Therefore, we compute $P(Y=y|do(X=x))$ (abbreviated as $P(y|do(x))$ later) as follows:
\begin{equation}\label{eq:backdooradj}
    P(Y=y|do(X=x)) = \sum_c P(Y=y|X=x,C=c)P(C=c)
\end{equation}

The above equation is based on the assumption that the confounder variable, which also satisfies the backdoor criterion, is measurable. However, in recommender systems, there may exist various unobserved or unmeasurable confounding variables (as mentioned in Section \ref{sec:intro}).
% (see Figure \ref{fig:adjustment}(b as an illustration)).
% Consider the case when variable $C$ is unobserved or unmeasurable (Figure \ref{fig:adjustment}(b)). 
To address this setting, we introduce the front-door criterion.

\begin{mydef}
(Front-door Criterion) \cite[p.69]{pearl2016causal} Given an ordered pair of variables $(X,Y)$ in a causal graph $\mathcal{G}$, a set of variables $Z$ satisfies the front-door criterion with respect to $(X,Y)$ if $Z$ satisfies the following conditions:
\begin{itemize}
\renewcommand\labelitemi{--}
    \item $Z$ intercepts all directed paths from $X$ to $Y$.
    \item There is no unblocked back-door path from $X$ to $Z$.
    \item $X$ blocks all back-door paths from $Z$ to $Y$.
\end{itemize}
\end{mydef}

Given a set of variables that satisfies the front-door criterion, we can identify the causal effect with unobserved confounders \cite{pearl2016causal}. 

\begin{mydef}
(Front-door Adjustment) \cite[p.69]{pearl2016causal} If a set of variables $Z$ satisfy the front-door criterion related to an ordered pair of variables $(X,Y)$, and if $P(x,z)>0$, then the causal effect of $X$ on $Y$ is identifiable and is given by
\begin{equation}\label{eq:frontdooradj}
    P(y|do(x)) = \sum_z P(z|x)\sum_{x^\prime}P(y|x^\prime, z) P(x^\prime)
\end{equation}
\end{mydef}

We take Figure \ref{fig:adjustment}(b) as an example. Although variable $C$ satisfies the back-door criterion, it is not measurable, so the back-door adjustment (Eq.\eqref{eq:backdooradj}) cannot be applied in this example. Although the back-door adjustment is not applicable here, given that variable $M$ satisfies the front-door criterion, we can use the front-door adjustment to handle unmeasurable or unobserved confounders. Intuitively, the desired effect can be expressed as follows
\begin{equation}\label{eq:frontpart}
    P(y|do(x)) = \sum_m P(m|do(x))P(y|do(m))
\end{equation}

Since the only parent node of variable $M$ is $X$ (thus $P(m|do(x))=P(m|x)$ \cite{pearl2016causal}), and variable $X$ satisfies the back-door criterion with respect to $(M,Y)$ so that we can apply back-door adjustment for $P(y|do(m))$, therefore, Eq.\eqref{eq:frontpart} can be further derived as
\begin{equation}\label{eq:frontdoorexample}
    % P(y|do(x)) = \sum_m P(m|do(x))\sum_{x^\prime}P(y|x^\prime, m)P(x^\prime)
    P(y|do(x)) = \sum_m P(m|x)\sum_{x^\prime}P(y|x^\prime, m)P(x^\prime)
\end{equation}

We can see Eq.\eqref{eq:frontdoorexample} is exactly the front-door adjustment as Eq.\eqref{eq:frontdooradj}.

\section{The Deconfounded Model}\label{sec:model}
In this section, we first introduce our designed causal graph depicting user behaviors with unobserved confounders. We then elaborate on our Deconfounded Causal Collaborative Filtering (DCCF) model. 
The primary notation used in this section is detailed in Table \ref{tab:notation}.

\begin{table}[]
    \centering
    \begin{tabular}{c p{10cm}}
    \toprule
        \textbf{Symbol} & \textbf{Definition} \\\midrule
        $U$, $u$ & The user variable and the corresponding specific values\\
        $V$, $v$ & The item variable and the corresponding specific values\\
        $M$, $m$ & The inherent item features and the corresponding specific values\\
        $Y$, $y$ & The preference variable and the corresponding specific values\\
        $C$ & The confounder variable\\
        $\mathcal{U}$, $\mathcal{V}$ & The user set and the item set\\
        $\mathbf{u}$,$\mathbf{v}$ & The representation of user $u$ and item $v$\\
        $\mathbf{U}$, $\mathbf{V}$ & The representation matrix of users and items\\
        $m_v$, $\mathbf{m}_v$ & The inherent item feature of item $v$ and the latent representation of the inherent item feature of item $v$\\
        $D_{uv}$ & The dimension of the latent representation of users and items\\
        $D_m$ & The dimension of the latent representation of inherent item features\\
        % $V_t$, $v_t$ & The variable denotes exposed item at time $t$ and the corresponding specific values\\
        % $X_t$, $x_t$ & The variable denotes user history at time $t$ and the corresponding specific values\\
        % $Y_t$, $y_t$ & The variable denotes the preference score at time $t$ and the corresponding specific values\\
        % $x^*_t$ & The observed user history at time $t$ in real world\\
        % $x'_t$ & The counterfactual user history at time $t$ which is unobserved in real world\\
        % $v^*_t$ & The observed exposed item at time $t$ in real world\\
        % $v'_t$ & The counterfactual exposed item at time $t$ which is unobserved in real world\\
        % $y^*_t$ & The observed preference score at time $t$ in real world\\
        % $y'_t$ & The counterfactual preference score at time $t$ which is unobserved in real world\\
        
         \bottomrule
    \end{tabular}
    \caption{Notations}
    \vspace{-20pt}
    \label{tab:notation}
\end{table}

\subsection{The Causal Graph}\label{sec:graph}

We formulate the process of user-system interaction as a causal graph shown in Figure \ref{fig:causal}. We explain the rationality of our designed causal graph from the view of data generation as follows:

\begin{itemize}[leftmargin=*]
    \item Edges $\{U,C\}\rightarrow V$ denote that the user will determine which item to interact with, and the unobserved confounder may affect user's decision.
    \item Edge $V \rightarrow M$ denotes that the item will determine the inherent item features. Here, the inherent features of the items that are designed as is from the beginning and will never change. For example, a silver macbook has a color as silver, brand as Apple, memory as 8G, OS as Macintosh. Thus the inherent item feature is solely determined by the item $V$ and will not be directly influences by user $U$ and confounders $C$.
    \item Edge $C \rightarrow U$ denotes that the unobserved confounders may affect user's preference. For example, as we exemplified in Section \ref{sec:intro}, weather, as a confounder, may personally affect users and lead to different preference.
    \item Edges $\{U,M,C\}\rightarrow Y$ denote that the user preference score is determined by user $U$, inherent item feature $M$ and unobserved confounders $C$. Specifically, the user's preference on a certain item may be affected by inherent item features (e.g., whether the functional attributes of the item conform to user preferences) and the confounding factors (e.g., the weather, the mood, etc.).
\end{itemize}

% First, the user will determine which recommended item(s) to interact with. Then, according to the corresponding inherent item features (i.e. the inherent features of the items that are designed as is from the beginning and will never change. For example, a silver macbook has a color as silver, brand as Apple, memory as 8G, OS as Macintosh.), user will reveal his or her preference feedback on the item. The whole process can be affected by confounders. Combining global confounders and personal confounders as introduced in Figure \ref{fig:confounderexample}, we design the causal graph as shown in Figure \ref{fig:causal}. In this causal graph, we are actually taking inherent item features as mediators between item node $V$ and preference node $Y$, i.e., $M$ is not directly influenced by user node $U$ and confounders $C$. 

The effect of global confounders (e.g., the air temperature as mentioned in Section \ref{sec:intro}) is represented by edge $C \rightarrow V$ and edge $C\rightarrow Y$. The effect of personal confounders (e.g. the mood as mentioned in Section \ref{sec:intro}) is represented by edge $C\rightarrow U$, edge $C \rightarrow V$ and edge $C\rightarrow Y$.

Setting variable of inherent item features as the mediator is intuitive and reasonable because inherent item features such as the color or brand of products and the genre or director of movies are inherent and fixed properties of an item, which are not influenced by users or external confounders. The inherent item features are not directly affected by users or confounders. Considering a general example, a user will click an item during browsing the website, where the inherent item features may not visible to users at this stage (for example, amazon homepage recommendations only show an image without any text information about the inherent item features). After clicking the item, based on the detailed inherent item features, combining with the effect of unobserved confounders and personal preference, the user will decide like or not, purchase or not, etc.
Our model aims to estimate the user's true preference of an item that is decided by the inherent item features and not affected by confounders.

In order to achieve our purpose during implementation, it is crucial to carefully select the inherent item features to serve as the mediator. Not all item features can be considered as inherent features since the mediator should not be directly affected by users or confounders. Specifically, we only select objective features, including features related to the appearance and functionality of the item, such as item description, color, and other similar features. Some other item features, such as sales rank and related products, are generated based on user's subjective reviews or user interactions and cannot be selected as the mediator. In real-world scenarios, it may not always be feasible for the selected mediator to completely intercept the causal pathways from the items to the preference variable, resulting in the possibility of direct effects from items to preference. Such a situation violates the front-door criterion and may have a negative impact on the deconfounding performance. However, if the mediator is chosen appropriately, it can capture a significant portion of the causal effect, making the direct effect weak and negligible in predicting preference. Hence, the impact on performance may not be significant. We will show some experimental results to illustrate it in Section \ref{sec:experiments}.
% Meanwhile, user's preference towards those inherent item features determines the true user-item preference which is not affected by confounders.

It is worth clarifying that the causal graph is used for describing how the collected data was generated. We ultimately build a recommendation model that can leverage this causal graph to produce better estimations of user preferences.

\subsection{Estimated Preference Score}\label{sec:preference}
\begin{figure}[t!]
    \centering
    \includegraphics[width=0.6\linewidth]{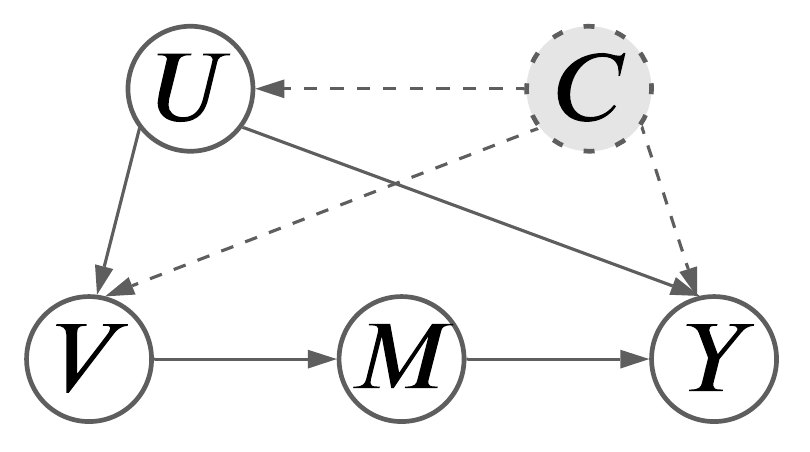}
    \vspace{-5pt}
    \caption{The meaning of variables are: $U$ represents user, $V$ represents exposed item, $M$ represents mediator (i.e., the inherent item features), $Y$ represents user's preference, and $C$ represents unobserved confounders. Dashed nodes represent unobserved variables and dashed arrows represent causal relations pointing from unobserved variables.}
    \label{fig:causal}
    \vspace{-5pt}
\end{figure}

In this section, we estimate user preference from a causal view, which aims to answer a \textit{what if} question: what would be the user’s preference on an item if we intervene to recommend the item to the user. In particular, we estimate a user's preference if a certain item is recommended to the user, which can be represented as $P(y|u,do(v))$. The estimation of the preference score $P(y|u,do(v))$ is based on the designed causal graph in Figure \ref{fig:causal}. From the causal graph as shown in Figure \ref{fig:causal}, it seems that the desired estimation can be calculated by the back-door adjustment \citep[p.61]{pearl2016causal} (as Eq.\eqref{eq:backdooradj}) since a set of variables $\{C,U\}$ satisfies the back-door. However, the back-door adjustment is not applicable because variable $C$ is unobserved or unmeasurable. Therefore, considering the existence of unobserved confounders, we apply front-door adjustment \citep[p.69]{pearl2016causal} (following Eq.\eqref{eq:frontdooradj}) to estimate user's preference on items. Concretely, We estimate the preference score of a $(u,v)$ pair as follows:
\begin{equation}\label{eq:frontdoorestimation}
    P(y|u,do(v)) = \sum_m P(m|v) \sum_{v^\prime} P(v^\prime|u)P(y|u, v^\prime, m)
\end{equation}
where $y$ represents the value of preference score and $m$ represents a specific value of inherent item features $M$. Here $m$ integrates all inherent features of an item. For example, an item has inherent features including color, brand, etc. $m$ represents a inherent feature profile which integrates the specific value of each inherent features, such as color as silver, brand as Apple, etc.

To utilize the item features into model prediction, we represent the item features into the continuous space, then we can rewrite the preference score as follows:
\begin{equation}\label{eq:frontdoor_int}
    P(y|u,do(v)) = \int_{\mathbf{m}} P(\mathbf{m}|v) \sum_{v^\prime} P(v^\prime|u)P(y|u, v^\prime, \mathbf{m}) d\mathbf{m}
\end{equation}
We denote the described feature of item $v$ as $\mathbf{m}_v$, and consider the conditional distribution of item feature as $P(\mathbf{m}|v) \sim \mathcal{N}(\mathbf{m}_v, \sigma_m)$. Using distribution instead of deterministic value is intuitive and reasonable in practice, because the feature of individual product may slightly differ from the described feature for reasons such as quality control issue. 

To calculate the integration over item's inherent feature, we apply Monte Carlo Sampling. More specifically, we sample $d$ values of $\mathbf{m}$ (i.e., $\mathbf{m}_1,\mathbf{m}_2,\cdots,\mathbf{m}_d$) based on the distribution of $P(\mathbf{m}|v)$. Therefore, we can convert the calculation in Eq.\eqref{eq:frontdoor_int} as follows:
\begin{equation}\label{eq:frontdoormcs}
    P(y|u,do(v)) = \frac{1}{d}\sum_{j=1}^d \sum_{v^\prime} P(v^\prime|u)P(y|u, v^\prime, \mathbf{m}_j)
\end{equation}

Real-world systems may include a large scale of items, therefore, it is unrealistic to involve all items to estimate the preference of a user-item pair. To tackle this issue, we modify the original front-door adjustment to a sample-based one that is more suitable and realistic for the recommendation scenario. Specifically, we randomly sample a set of items while calculating $P(y|u,do(v))$ as a trade-off between efficiency and deconfounding performance (more discussion and results are provided in Section \ref{sec:expomodels}). The item set for a user-item pair $(u,v)$ is represented as $R(u,v)$ (including sampled items and item $v$), and the number of sampled items is $n$, which is a hyper-parameter to be selected. 
% It is worth noting that we include item $v$ because, in the training process, item $v$ is observed in the dataset, thus including item $v$ may be helpful for recommendation performance.
We can rewrite Eq.\eqref{eq:frontdoormcs} into a sample-based format.
\begin{equation}\label{eq:samplefrontdoor}
    P(y|u,do(v)) = \frac{1}{d}\sum_{j=1}^d\sum_{v^\prime\in R(u,v)} P(v^\prime|u)P(y|u, v^\prime, \mathbf{m}_j)
\end{equation}
For each item $v^\prime \in R(u,v)$ in Eq.\eqref{eq:samplefrontdoor}, its exposure probability (i.e. $P(v^\prime|u)$) and conditional preference (i.e. $P(y|u, v^\prime, \mathbf{m}_j)$) are two required values. We will introduce how to calculate them respectively.

\begin{figure*}
    \centering
    \includegraphics[width=\linewidth]{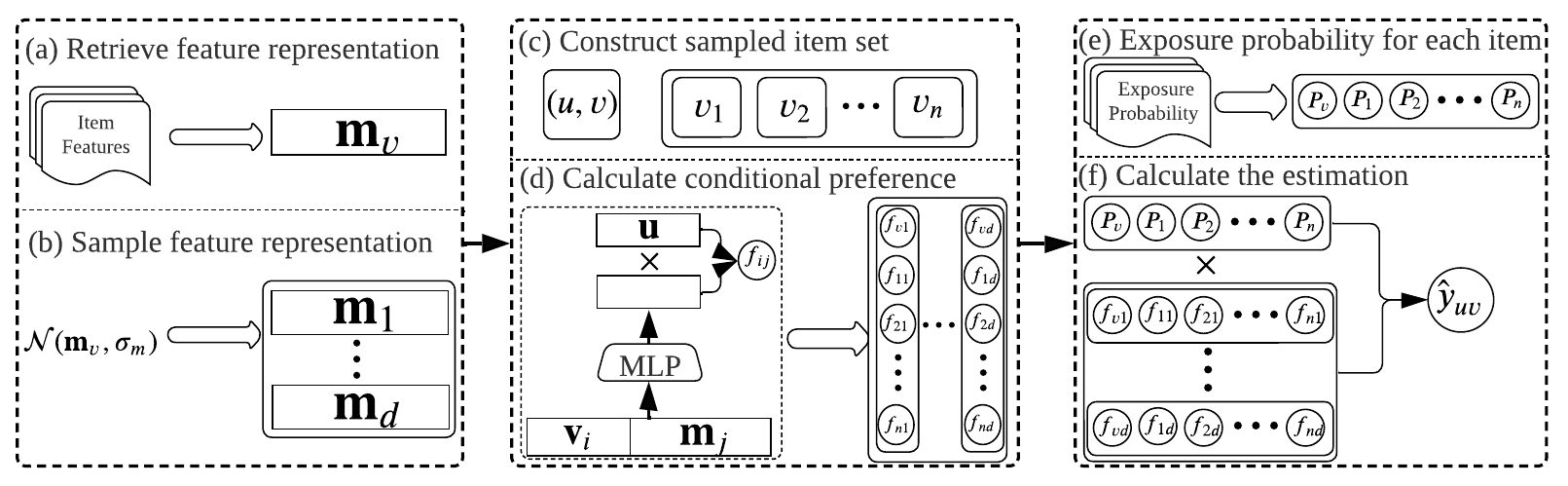}
    \vspace{-15pt}
    \caption{The process of estimating preference $\hat{y}_{uv}$. $\mathbf{m}_v$ is the feature representation for item $v$ described in the dataset; $\mathbf{m}_1 \cdots \mathbf{m}_d$ are sampled inherent item feature from $\mathcal{N}(\mathbf{m}_v,\sigma_m)$; $v_1,v_2\cdots v_n$ are $n$ sampled items; $f_{ij}$ is the conditional preference $f(u,v_i,m_j)$ of user $u$ on item $v_i$ with inherent item feature $\mathbf{m}_j$ (Eq.\eqref{eq:conditionalpref}); $P_i$ is the exposure probability of item $v_i$ on user $u$ (i.e., the $P^e_{uv_i}$ value according to Eq.\eqref{eq:ips_expo}); $\hat{y}_{uv}$ is the final preference score between user $u$ and item $v$ (Eq.\eqref{eq:finalestimation}).
    % Each green circle represents the conditional preference score of one item in $R(u,v)$, each blue circle represents the exposure probability of one item in $R(u,v)$. The calculation of conditional preference score for each item is shown in (c) which corresponds to Eq.\eqref{eq:conditionalpref}. The exposure probability for each item is obtained from the exposure model. The final user-item preference $\hat{y}_{uv}$ is the sum over the products between exposure probability and conditional preference for each item in the sampled item set.
    }
    \label{fig:model}
    \vspace{-5pt}
\end{figure*}

\subsubsection{\textbf{Calculate the Exposure Probability}}

The casual graph and front-door adjustment enable us to expand the expression of $P(y|u,do(v))$ (as Eq.\eqref{eq:samplefrontdoor}). However, to obtain the value of $P(y|u,do(v))$ for recommendation, we need to calculate the value of $P(v’|u)$ as exposure probability which is a part of the expanded $P(y|u,do(v))$ expression. The exposure probability $P(v|u)$ represents the probability of an item $v$ being exposed to a certain user $u$.
% The specific probabilities are available in the dataset, however, 
The implicit feedback in the dataset contains binary interactions, which indicates which user interacted with which item.
% \yz{could be problematic to use interaction for calculating exposure, e.g., some items are exposed but user didn't click, some other items that user likes are not exposed}.
If we take the interacted items as exposed (positive) samples and non-interacted items as unexposed (negative) samples (we will relax this assumption in the following), then we can train an exposure model based on pair-wise learning using the following objective:
\begin{equation}\label{eq:exposure}
    \min -\sum_{u\in \mathcal{U}}\sum_{v_u^+}\sum_{v_u^-}\ln(\sigma(P^e_{uv_u^+} - P^e_{uv_u^-}))
\end{equation}
where $\sigma(\cdot)$ is the sigmoid function: $\sigma(x) = \frac{1}{1+e^{-x}}$, $v_u^+$ and $v_u^-$ are positive and negative items for user $u$ respectively, $P^e_{uv}$ is the predicted exposure probability for user-item pair $(u,v)$, which corresponds to $P(v|u)$ in Eq.\eqref{eq:samplefrontdoor}. More specifically, we can apply Matrix Factorization (MF) \cite{koren2009matrix} to calculate $P^e_{uv}$:
% which considers user, item and global offset terms for matrix factorization.
\begin{equation}\label{eq:biasedMF}
P^e_{uv} = \mathbf{p}_u\mathbf{q}_v^T+b_u+b_v+b_g,
\end{equation}
where $\mathbf{p}_u$ and $\mathbf{q}_v$ are embedding representations for user $u$ and item $v$, $b_u$ is the user preference offset term, $b_v$ is item preference offset term, and $b_g$ is global offset term. $\mathbf{p}_u$, $\mathbf{q}_v$, $b_u$, $b_v$, and $b_g$ are all learned in training.

Although we can obtain exposure probabilities based on Eq.\eqref{eq:biasedMF}, it ignores the bias in the observational data, that the non-interacted items may not always represent non-exposed samples. Therefore, the exposure probabilities for some non-interacted items can be underestimated, which leads to inaccurate causal estimation (i.e., Eq.\eqref{eq:samplefrontdoor}) and further sacrifices the performance. 

To address this issue and obtain more accurate exposure probabilities, some efforts have been made, such as doubly robust model \cite{wang2019doubly} and predictive model \cite{marlin2009collaborative}. For simplicity, we apply the simple IPS-based debiasing technique \cite{saito2020unbiased}.
% \yz{add citation}. 
Concretely, we use propensity scores to correct the predicted probability.
\begin{equation}\label{eq:ips_expo}
    P^e_{uv} = (\mathbf{p}_u\mathbf{q}_v^T+b_u+b_v+b_g) / p_{uv}
\end{equation}
where $p_{uv}$ is the propensity score for user-item pair $(u,v)$ (calculated as Eq.\eqref{eq:propensity}). Theoretically, Eq.\eqref{eq:ips_expo} will return more accurate and unbiased exposure probabilities \cite{yang2018unbiased}.
% \yz{add citation}. 
We will discuss the empirical influence of different exposure models in the experiments section (Section \ref{sec:expomodels}).

% For each user-item pair $(u,v)$, we have user representation $\mathbf{u}$ for user $u$ and item representation $\mathbf{v}$ for item $v$. In particular, we assume that the exposure probability is proportional to the dot product of user and item representations (i.e. $P(v|u)\propto \mathbf{uv}^T$). Inspired by \cite{}, we use user and item representation to estimate $P(v^\prime|u)$ in Eq.\eqref{eq:samplefrontdoor}.
% \begin{equation}\label{eq:expoprob}
%     P(v|u) = \frac{exp(\mathbf{uv}^T)}{\sum_{v^\prime\in R(u,v)} exp(\mathbf{uv^\prime}^T)}
% \end{equation}

\subsubsection{\textbf{Calculate the Conditional Preference}}

To estimate the desired value of $P(y|u,do(v))$, the conditional preference $P(y|u,v^\prime, \mathbf{m}_j)$ is an essential component. Recall that $u$ represents a certain user, $v^\prime$ represents a specific item, and $\mathbf{m}_j$ represents the representation of inherent item feature. We design a function $f:U,V,M\rightarrow Y$ which takes users, items and inherent item features as input and returns preference scores such that $P(y|u,v^\prime, \mathbf{m}_j)$ is proportional to the corresponding conditional preference score.
\begin{equation}\label{eq:propto}
    P(y|u,v^\prime, \mathbf{m}_j) \propto f(u,v^\prime, \mathbf{m}_j)
\end{equation}
For a triple $(u,v^\prime,\mathbf{m}_j)$, we first use a
Multi-Layer Perceptron (MLP) to encode the item embedding and latent representation of inherent item features into a vector, and then use the dot product between the obtained vector and the user embedding to obtain the preference score.
\begin{equation}\label{eq:conditionalpref}
    f(u,v^\prime,\mathbf{m}) = \mathbf{u}\cdot \phi(\mathbf{W}_\ell\phi( \mathbf{W}_{\ell-1} \phi (\dots \phi(\mathbf{W}_1
    \begin{bmatrix}
    \mathbf{v}^\prime\\
    \mathbf{m}_j
    \end{bmatrix})\dots)))^T
\end{equation}
where $\mathbf{u},\mathbf{v}^\prime\in \mathbb{R}^{D_{uv}}$ are the user and item latent embedding vectors in $D_{uv}$-dimensional space (and it is not same as $\mathbf{p}_u$ and $\mathbf{q}_v$ in Eq.\eqref{eq:ips_expo} since $\mathbf{p}_u$ and $\mathbf{q}_v$ are used to calculate the exposure probability while $\mathbf{u}$ and $\mathbf{v}^\prime$ are used to calculate the deconfounded preference); $\mathbf{m}_j\in \mathbb{R}^{D_m}$ is the representation of inherent item features (i.e., latent representation that integrates all inherent features of item $v$) in $D_m$-dimensional space ($D_{uv}$ and $D_m$ are not necessarily equal), which is retrieved from data; $\ell$ is the number of layers in MLP; $\mathbf{W}_i, i=1,\dots, \ell$ are weight matrices to be learned; and $\phi(\cdot)$ is the rectified linear unit (ReLU) activation function: $\phi(x)=\max(0,x)$.

\subsubsection{\textbf{Calculate the Expectation}}
Combined Eq.\eqref{eq:samplefrontdoor} with Eq.\eqref{eq:propto}, we have our estimation as follows:
\begin{equation}\label{eq:finalestimation}
    P(y|u,do(v)) = \frac{1}{d}\sum_{j=1}^d\sum_{v^\prime\in R(u,v)} P(v^\prime|u)f(u, v^\prime, \mathbf{m}_j)
\end{equation}
where $P(v^\prime|u)$ is obtained by Eq.\eqref{eq:ips_expo} and $f(u, v^\prime, m_v)$ is calculated by Eq.\eqref{eq:conditionalpref}. The calculation of $\hat{y}_{uv}$ is shown in Figure \ref{fig:model}.

\setlength{\textfloatsep}{.1cm}% Remove \textfloatsep
\begin{algorithm}[t]
    \caption{\textbf{D}econfounded \textbf{C}ausal \textbf{C}ollaborative \textbf{F}iltering (\textbf{DCCF})}
    \begin{flushleft}
    \textbf{Input:} \\
    Observed interactions as user-item pair $(u,v)$ in the training data;\\
    The inherent item feature representation $\mathbf{m}$ for all items in $D_m$-dimensional space;\\
    L2-norm regularization weight: $\lambda_\Theta$;\\
    The number of sampled items for each user-item pair: $n$\\
    \textbf{Output:} \\
    user representations $\mathbf{U}$; item representations $\mathbf{V}$; MLP network $\mathbf{W}_i$
    \end{flushleft}
    \begin{algorithmic}[1]
        \STATE Train the exposure model and obtain $P(v|u)$ as Eq.\eqref{eq:ips_expo}
        % \STATE Calculate all exposure probabilites $P(v|u)$ as Eq.\eqref{eq:ips_expo}
        \STATE Random initialization of $\mathbf{U}$, $\mathbf{V}$, $\mathbf{W}_i$
        \WHILE{\textit{not converged}}
        \FOR{\textit{each batch in training set}}
        \STATE $Loss\leftarrow 0$
        \FOR{\textit{each training sample $(u,v)$ in the batch}}
        \STATE $R(u,v) \leftarrow$ $v~\cup$ Randomly sample $n$ items
        \STATE $\mathbf{m}_j|_{j=1}^d \leftarrow$ Sample inherent item features $d$ times following distribution $\mathcal{N}(\mathbf{m}_v, \sigma_m)$
        % \STATE $R(u,v) \leftarrow R(u,v)\cup \{v\}$
        \STATE $\hat{y}_{uv} \leftarrow \frac{1}{d}\sum_{j=1}^d\sum_{v^\prime\in R(u,v)} P(v^\prime|u)f(u, v^\prime, \mathbf{m}_j)$
        \STATE $v^- \leftarrow$ Obtain negative sample
        \STATE $R(u,v^-) \leftarrow$ $v^-~\cup$ Randomly sample $n$ items
        \STATE $\mathbf{m}'_j|_{j=1}^d \leftarrow$ Sample inherent item features $d$ times following distribution $\mathcal{N}(\mathbf{m}_{v^-}, \sigma_m)$
        % \STATE $R(u,v^-) \leftarrow R(u,v^-)\cup \{v^-\}$
        \STATE $\hat{y}_{uv^-} \leftarrow \frac{1}{d}\sum_{j=1}^d\sum_{v^\prime\in R(u,v^-)} P(v^\prime|u)f(u, v^\prime, \mathbf{m}'_j)$
        \STATE $Loss\leftarrow Loss + \ln(\sigma(\hat{y}_{uv} - \hat{y}_{uv^-}))$
        \ENDFOR
        \STATE $Loss\leftarrow Loss + \lambda_\Theta \left\Vert\Theta\right\Vert_2^2$
        \STATE Update $\mathbf{U}$, $\mathbf{V}$, $\mathbf{W}_i$
        \ENDFOR
        \ENDWHILE
        \RETURN $\mathbf{U}$, $\mathbf{V}$, $\mathbf{W}_i$
    \end{algorithmic}
    \label{alg:model}
\end{algorithm}

\subsection{Model Learning}
In this section, we will introduce details about model learning, i.e., how to learn the estimation as Eq.\eqref{eq:finalestimation}. In this work, we use the pair-wise learning-to-rank algorithm \cite{rendle2012bpr} for training. Suppose we observe user $u$ interacted with item $v_i$ in the dataset, The estimated preference $\hat{y}^+_{ui}$ is obtained from Eq.\eqref{eq:finalestimation}. We sample another item $v_j\in \mathcal{V}$ that user $u$ did not interact with as a negative sample. The estimated preference for negative sample $v_j$ is represented as $\hat{y}^-_{uj}$. The difference between two estimated preferences is noted as $\hat{y}_{uij}$.
\begin{equation}
    \hat{y}_{uij} = \hat{y}^+_{ui} - \hat{y}^-_{uj}
\end{equation}

Given the observed interactions, we randomly sample a negative item for each user-item pair in implementation. Specifically, we use $\mathcal{V}^+_u$ to represent the observed items set for user $u$ ($v_i\in \mathcal{V}^+_u$) and $\mathcal{V}^-_u$ to represent the sampled negative item set for user $u$ ($v_j\in \mathcal{V}^-_u$). The recommendation loss function can be written as:
\begin{equation}\label{eq:recloss}
    \mathcal{L} = -\sum_{u\in \mathcal{U}}\sum_{v_i\in \mathcal{V}^+_u}\sum_{v_j\in\mathcal{V}^-_u}\ln(\sigma(\hat{y}_{uij}))+\lambda_\Theta \left\Vert\Theta\right\Vert_2^2
\end{equation}
where $\Theta$ represents all parameters of the model; $\lambda_\Theta$ is the L2-norm regularization weight; $\sigma(\cdot)$ is the sigmoid function: $\sigma(x) = \frac{1}{1+e^{-x}}$. It is worth mentioning that the representation of inherent item feature $\mathbf{m}$ and exposure probability model to calculate $P(v^\prime|u)$ are pre-trained and fixed during the training of our model. The overall algorithm is provided in Algorithm \ref{alg:model}.

% By now, we learn user, item representations and corresponding network weights to optimize the loss shown in Eq.\eqref{eq:recloss}. However, we cannot guarantee that the dot product of user and item representation is actually proportional to the exposure probability. Therefore, to make sure that the representations satisfy our assumed physical meaning, similar to \cite{chen2020neural}, we add a regularizer to constraint their behavior. To achieve this, we assume that interacted item has higher exposure probability than uninteracted item. Therefore, we define a exposure loss.
% \begin{equation}\label{eq:exposureloss}
%     \mathcal{L}_e = -\sum_{u\in \mathcal{U}}\sum_{v_i\in \mathcal{V}^+_u}\sum_{v_j\in\mathcal{V}^-_u}ln(\sigma(\mathbf{uv}_i^T-\mathbf{uv}_j^T))
% \end{equation}
% Now we integrate the exposure loss into the recommendation loss to get the final loss function.
% \begin{equation}\label{eq:finalloss}
%     \mathcal{L} = \mathcal{L}_r + \lambda_e \mathcal{L}_e
% \end{equation}
% where $\lambda_e$ is the weight for the exposure loss, which is a hyper-parameter to be tuned.

% \iffalse
\subsection{Identifiability Issue}

In this section, we discuss the identifiability of our estimated causal preference.
Generally speaking, identifiability in causal inference indicates whether we can use observed data to estimate target values with consistency \cite{pearl2016causal,mohan2021graphical}. From a view of causal inference in statistics, $(y,u,v',\mathbf{m}_i)$ is unobserved in the data thus the target value $P(y|u,do(v))$ (Eq.\eqref{eq:finalestimation}) is unidentifiable. Such non-identifiability problems are common problems in recommender systems. The reason is that recommender systems are often estimated from datasets containing a very high portion of missing data. Many works \cite{marlin2011recommender,wang2019doubly,saito2020unbiased} have shown that the missing data is often missed not at random (MNAR) in reality, and recent research \cite{mohan2021graphical} has shown that the target value is unidentifiable to any method if MNAR exists. 
Even if the target values are unidentifiable, in many practical intelligent systems such as recommender systems, we still need to estimate reasonably good values for such unidentifiable variables \cite{zhang2021causal,yang2021top,wang2021deconfounding}, so as to make it possible to provide practical services for users. 
One example is to estimate $P(y|u,v’)$ where $v’$ is an item that user u never interacted with before. Real-world recommender systems usually have billions of items, but each individual user only has interacted with tens or at most hundreds of them. Even though most of the $(u,v’)$ interactions cannot be observed, we still need to estimate the probability of such interactions, because only by doing so can we provide recommendations to users \cite{zhang2021causal,yang2021top,wang2021deconfounding}. If we give up any attempt to estimate such unidentifiable variables, then many intelligent services will be impossible in practice.

In our model, the key to estimate such unidentifiable variables is through “collaborative learning”. 
The basic idea is that, although the target value is unobserved, there could exist some other similar examples whose values are observed, so that we can borrow their values to estimate the target value. By doing so, the examples can “collaborate” with each other so as to help estimate the values of each other, which is the key idea of “collaborative filtering” – one type of collaborative learning techniques. 
In our model, 
More specifically, given $(y, u, v, \mathbf{m}_v)$ is observed, $(y, u, v’, \mathbf{m}_j)$ is indeed unobserved, but if $v$ and $v’$ share similar features, $\mathbf{m}_v$ and $\mathbf{m}_{v’}$ will also be similar representations learned by language models. 
Specifically, when we calculate $P(y|u,v’,\mathbf{m}_j)$, considering $\mathbf{m}_j$ is similar to $\mathbf{m}_v$ (i.e., $\mathbf{m}_j$ is sampled from $\mathcal{N}(\mathbf{m}_v,\sigma_m)$), if $v’$ shares similar inherent item feature with $v$, then $\mathbf{m}_{v’}$ is also similar to $\mathbf{m}_j$, thus $P(y|u,v’,\mathbf{m}_j)$ is similar to observed $P(y|u,v,\mathbf{m}_v)$. On the contrary, if $v’$ is very different from $v$, then $\mathbf{m}_{v’}$ is also different from $\mathbf{m}_j$, which will lead to $P(y|u,v’, \mathbf{m}_j)$ being quite different from $P(y|u,v,\mathbf{m}_v)$. The above is a direct implication of the collaborative learning principle in representation learning techniques.
% \fi

\section{Experiments}\label{sec:experiments}

In this section, we conduct experiments to demonstrate the efficacy of our deconfounded recommender (DCCF) on real datasets. In particular,  we aim to answer the following research questions:

\begin{itemize}
    \item \textbf{RQ1}: How do deconfounded models perform on real-world datasets?
    \item \textbf{RQ2}: How does DCCF perform compared with other deconfounded models?
    \item \textbf{RQ3}: How does the number of sampled items influence the performance and efficiency?
    \item \textbf{RQ4}: How different exposure models influence the performance?
    
\end{itemize}

We will first describe datasets, baselines and implementation details and then provide our results and analysis.

\begin{table}[t]
    \centering
    % \begin{adjustbox}{width=0.95\linewidth}
    \begin{tabular}{l|c|c|c|c}
    \toprule
        Dataset & \# users & \# items & \# interactions & Sparsity\\\midrule
        Electronics & 33602 & 16448 & 788143 & 99.86\%\\
        CDs and Vinyl  & 6867 & 6953 & 249456 & 99.48\%\\
        Yelp & 21439 & 15914 & 996118 & 99.71\%\\
        % Grocery and Gourmet Food& 1645 & 1120 & 38154 \\
        \bottomrule
    \end{tabular}
    % \end{adjustbox}
    \caption{Statistics about datasets}
    \label{tab:datasets}
    % \vspace{-10pt}
\end{table}

\subsection{Data Description}
\subsubsection{\textbf{Real-world Data}}

Our experiments are conducted on the Amazon Review Datasets\footnote{https://nijianmo.github.io/amazon/index.html} \cite{ni2019justifying} and Yelp dataset \footnote{https://www.yelp.com/dataset}. Amazon Review Datasets include user, item, rating and product metadata spanning from May 1996 to October 2018. More specifically, we experiment with two categories in Amazon Review Datasets, (1) \textit{Electronics} and (2) \textit{CDs and Vinyl}. Yelp dataset includes user, business, rating and business information. For each dataset, we consider ratings $\geq 4$ as positive feedback (likes) and ratings $\leq 3$ as negative feedback (dislikes). We sample 70\% of positive interaction as training data, 10\% as validation data and the remaining 20\% as testing data. The statistics about the datasets are shown in Table \ref{tab:datasets}.

Some other benchmark datasets such as Movielens datasets contains the genre information as inherent feature, however, this is a categorical feature and many movies are of the same genre, and thus it is difficult to distinguish different movies using such information. Therefore, Movielens datasets are not suitable for our setting. Yahoo! R3\footnote{\url{https://webscope.sandbox.yahoo.com/catalog.php?datatype=r}} and Coat Shopping \cite{schnabel2016recommendations} are suitable datasets for evaluating unbiased or deconfounded methods since both two datasets have asked users to rate randomly exposed items. However, in our front-door adjusted method, the inherent item features (cannot be extracted from reviews or interactions) are required to estimate the desired causal preference, and the inherent item features are extracted from image or text information. Yahoo R3 only contains user interaction and rating information and Coat only contains categorical feature (similar to Movielens), and thus Yahoo!R3 and Coat are not suitable for our experiments.

Although the above two datasets (Yahoo! R3 and Coat Shopping) for evaluating deconfounded models are not suitable for our setting, we apply a commonly used splitting strategy to evaluate deconfounded models. Following \cite{bonner2018causal}, we apply the skewed splitting strategy (denoted as SKEW) for all datasets, which is commonly used strategy for evaluating unbiased or deconfounded model. It simulates the results of randomized experiment by exposing as uniformly as possible each user to each item in testing set. Concretely, we first calculate the popularity of each item and find the least popular item. For all other items, We use its popularity ratio compared with the least popular item to calculate the probability of being divided into the test set. For example, a 100 times more popular item than the least popular item will have a probability of being sampled in the test set as 1\%.
To avoid from the least popular item being not available in training, similar to \cite{bonner2018causal}, we set the maximum probability for a user-item interaction record to be put into the test set as 0.9. The SKEW dataset is used to test if our model can perform well on randomized experiment results. Additionally, we also apply the traditional random splitting strategy (denoted as RAND), which considers each data sample with equal probability to be chosen into the testing set. All data and source code will be released when paper is published.

\subsubsection{\textbf{Synthetic Data}} The main purposes of using synthetic data are two fold: 1) to show the effectiveness of our model on mitigating the effect of unobserved confounders because the effect of unobserved confounders may not be measured in real-world data; 2) to observe the performance if the mediator cannot ``interpret'' all causal effect from item to preference. We generate two synthetic datasets \textbf{SD1} and \textbf{SD2} for the two purposes, respectively. For \textbf{SD1}, we follow the data generation process described in our designed causal graph (i.e., Figure \ref{fig:causal}). More specifically, we use a Matrix Factorization model to generate interactions, and use a neural network which takes users and inherent item features as input to obtain the feedback. For \textbf{SD2}, we combine two neural networks, one takes users and inherent item features as inputs to obtain the feedback, and the other takes users and items as inputs to obtain the feedback. In order to simulate a situation where the selected mediator captures most of the causal effects from items to preference, we scale down the output of the second neural network by a small value (e.g., 0.1). For both synthetic datasets, we assume that there are 900 users and 1000 items. The inherent item feature is generated by a neural network. When generating the data for training, we include the effect of confounders (randomly generated). When generating the data for testing, we eliminate the effect of confounders to obtain an unbiased testing set. For each user, we generate 20 interactions for training and 5 interactions for testing.

% Specifically, for each dataset, we apply two splitting strategies. One is the traditional random splitting strategy (denoted as RAND), which is commonly used in traditional recommendation scenario. It considers each data sample with equal probability to be chosen into the testing set.
% The other is skewed splitting strategy (denoted as SKEW) following \cite{bonner2018causal}, which is commonly used for evaluating unbiased or deconfounded models.  

\subsection{Baselines}
% To show the effectiveness of our proposed model, we compare it with some existing models. 
We introduce the baselines used in our experiments as follows. 
% In the following baselines, we include classical recommendation models, deconfounded recommendation models, and a variant of our model.

\begin{itemize}
    \item \textbf{MF} \cite{rendle2012bpr}: The matrix factorization model, which uses matrix factorization as the prediction function under the Bayesian personalized ranking framework. 
    \item \textbf{DMF} \cite{xue2017deep}: Deep Matrix Factorization is a deep learning model for recommendation, which filters the user-item interaction matrix through a neural network architecture for prediction.
    \item \textbf{DecRS} \cite{wang2021deconfounding}: A deconfounded recommendation model for alleviating bias amplification, which identifies the confounder as user historical distribution over item groups, and further applies back-door adjustment to mitigate the impact of the confounder.
    \item \textbf{IPW-MF} \cite{schnabel2016recommendations}: Inverse Propensity Weighting Matrix Factorization for recommendation, which adapts causal inference (inverse propensity weighting, specifically) on the matrix factorization model to handle the selection bias in observational data.
    \item \textbf{DCF} \cite{wang2020causal}: A deconfounded recommendation model,
    % applied on Matrix Factorization, 
    which uses an exposure model to construct a substitute confounder and then conditions on the substitute confounder for modeling.
    % to model the ratings
    \item \textbf{HCR} \cite{zhu2022mitigating}: A deconfounded recommendation model, which considers click behavior as mediator and designs a multi-task learning framework that simultaneously learns the probability of click and post-click behavior to mitigate confounding effect.
    \item \textbf{DCCF$\_$NS}: This is a no-item-sampling variant of our model. Specifically, for a user-item pair $(u,v)$, this variant only uses $P(v|u)f(u,v,m)$ in Eq.\eqref{eq:conditionalpref} as the estimated preference. It can be considered as a reweighting method.
    \item \textbf{DCCF$\_$ND}: This is a no-deconfounding (ND) variant of our model. Specifically, for a user-item pair $(u,v)$, this variant only uses $f(u,v,m)$ in Eq.\eqref{eq:conditionalpref} as the estimated preference. It can be considered as our model (DCCF) without the front-door adjustment, which loses the deconfounding ability.
\end{itemize}
The baselines include both classical association-based recommendation models (MF and DMF) and state-of-the-art deconfounded recommendation models (DecRS, IPW-MF, DCF and HCR), as well as two ablated variants of our own model (DCCF$\_$NS and DCCF$\_$ND). 
Many recent deconfounded methods such as DecRS mainly focus on mitigating one specific confounder, while our model is used to mitigate the effect of unobserved confounders.
% Since our paper is under a different problem setting from those previous works, we choose only DecRS as a baseline from recent deconfounded methods targeting one specific confounder. 
Some other deconfounded methods require certain user information such as user social networks \cite{gao2021deconfounding}, which is not suitable as a baseline for our problem setting. Finally, we use \textbf{DCCF} to denote the complete version of our model.

\subsection{Evaluation Metrics}
Our model aims to estimate the accurate preference with the existence of unobserved confounders. To evaluate the deconfounding performance, some deconfounded models, which mainly focus on one specific confounder, group items or users by the value of the confounder and compare the performance among groups. For example, \citet{zhang2021causal} treat item popularity as the confounder, thus the deconfounding performance is evaluated by grouping items by item popularity and calculating the recommendation ratio for each group. In our case, however, it is impossible to group users or items by unobserved confounders.

Since the deconfounded recommendation will obtain more accurate estimation of preference to improve the recommendation performance \cite{wang2020causal}, we evaluate our model and baselines on the top-$K$ recommendation task. We adopt three widely used metrics to evaluate recommendation performance including: \textit{Normalized Discounted Cumulative Gain@$K$} (nDCG@K for short), \textit{Recall@K} (Rec@K for short), and \textit{Precision@K} (Pre@K for short). The calculation of the three metrics is given as follows:
\begin{equation}
\begin{aligned}
    nDCG@K &= \frac{1}{|\mathcal{U}|}\sum_{u\in\mathcal{U}}\frac{DCG_u@K}{IDCG_u@K}\\
    Rec@K &= \frac{1}{|\mathcal{U}|}\sum_{u\in\mathcal{U}}\frac{|TP_u|}{|TP_u|+|FN_u|}\\
    Pre@K &= \frac{1}{|\mathcal{U}|}\sum_{u\in\mathcal{U}}\frac{|TP_u|}{|TP_u|+|FP_u|}
\end{aligned}
\end{equation}
where $DCG_u@K$ is the Discounted Cumulative Gain at position $K$ for user $u$, $IDCG_u@K$ is the Ideal Discounted Cumulative Gain through position $K$ for user $u$, $|TP_u|$ is the number of recommended items that are relevant for user $u$ at position $K$, $|TP_u|+|FN_u|$ is the total number of relevant items for user $u$, and $|TP_u|+|FP_u|$ is the total number of recommended items at position $K$ for user $u$. In our experiment, we refer to \cite{wang2020causal} and set $K=5$.

\subsection{Implementation Details}
We use the same training, validation and testing sets for all models, including our models and baseline models. 
% We evaluate models on the top-$K$ recommendation task. More specifically, models are evaluated based on nDCG@5, Recall@5, and Precision@5 Metrics. 
For evaluation, we apply \textit{real-plus-N} \cite{said2014comparative} to calculate the values of each metric. Concretely, for each user in the validation and testing set, we randomly sample 1000 negative items for ranking evaluation in real-world data and 100 negative items for synthetic data. All recommendation models adopt the Bayesian Personalized Ranking (BPR) \cite{rendle2012bpr} framework for pair-wise learning. During the training, we sample one negative item that the user did not interact with for each interacted user-item pair. We optimize the models using mini-batch Adam with a batch size of 128. 

% To get the feature representation of items, we apply a pre-trained sentence embedding model \footnote{we apply the pre-trained paraphrase-distilroberta-base-v1 sentence model in a public sentence transformer implementation: https://github.com/UKPLab/sentence-transformers} to represent the text information of an item in the meta-data into an embedding with dimension 768. More specifically, we convert the text information about title, feature, and description into embeddings separately, and take the average as the feature representation of the item.

To obtain the inherent feature representation of the items (i.e., $\mathbf{m}_v$ in Eq.\eqref{eq:conditionalpref}), we apply a pre-trained transformer-based sentence embedding model\footnote{we apply the pre-trained paraphrase-distilroberta-base-v1 sentence embedding model in a public transformer implementation: https://github.com/UKPLab/sentence-transformers} to represent the textual information of an item into a dense vector embedding with dimension 768. For the Amazon Review Datasets, the textual information comes from the 'title', 'feature', and 'literal description' of each item, which are included in the meta-data of the amazon review dataset. For the Yelp dataset, we include the 'name', 'address', 'city', 'state', 'attributes', and 'categories' of the business into the inherent feature, which can be found in the business information in the Yelp dataset.
We use the above pre-trained sentence transformer to convert the textual information into separate embeddings and further take the average as the inherent feature representation of the item.

For the hyper-parameters, we perform the grid-search and search the user/item embedding dimension from \{32,64,128\}, the structure of the neural network is a two-layer MLP with dimension 64 for deep models. For all recommendation models, we search the learning rate from \{0.0005,0.001,0.003,0.005\}, the $\ell_2$-regularization weight is chosen from \{1e-3, 1e-4, 1e-5\}. The total number of epoches is set to 100. For a fair comparison, we tune the parameters to the best performance for each model on the validation data based on nDCG@5. 
For our model, we sample 20 items for each user-item pair for calculating the expectation (i.e., the number of sampled items $n$ in Section \ref{sec:preference}) and sample 2 inherent item feature values for calculating the integration (i.e., the number of sampled item features $d$ in Section \ref{sec:preference}). We set the variance of the distribution of inherent item feature as 0.1 (i.e., $\sigma_m$ in Section \ref{sec:preference}). 
To obtain the exposure probability calculated based on Eq.\eqref{eq:ips_expo}, we estimate the user independent propensity scores as \cite{saito2020unbiased}:
\begin{equation}\label{eq:propensity}
    p_{*v}=\bigg(\frac{\sum_{u\in\mathcal{U}}y_{uv}}{\max_{v^\prime\in\mathcal{V}}\sum_{u\in\mathcal{U}}y_{uv^\prime}}\bigg)^\eta
\end{equation}
Here $y_{uv}$ is an indicator: $y_{uv}=1$ if the user-item pair $(u,v)$ is observed in the dataset, $y_{uv}=0$ otherwise. We set $\eta=0.5$ in the experiment. The running time of our model is about 40 seconds/epoch for \textit{Electronics}, 15 seconds/epoch for \textit{CDs and Vinyl} and 35 seconds/epoch for \textit{Yelp}.

\subsection{Results and Discussion}

In this section, we aim to answer \textbf{RQ1} and \textbf{RQ2}. For real-world data, the recommendation performance for models on the SKEW splitting datasets are shown in Table \ref{tab:SKEW_results}.
% The recommendation performance on baselines and our model are shown in Table \ref{tab:SKEW_results} and Table . We evaluate both \textit{Electronics} and \textit{CDs and Vinyl} datasets with the RAND and SKEW splitting, respectively. 
% The results include ranking metrics nDCG, recall and precision.
% For RAND and SKEW splitting, we strictly follow the 7:1:2 ratio to split train, validation and test data for each user, which makes the testing data the same size for each user on both splitting strategies. 
The SKEW splitting strategy simulates a test set as a result of randomized experiment, and the test distribution is different from the training distribution. The recommendation performance on the RAND splitting datasets are shown in Table \ref{tab:RAND_results}. The RAND splitting strategy is the regular splitting strategy that randomly splits data samples into testing set, thus the training and testing distributions are constant. Comparing the RAND and SKEW splitting strategies, we can observe that the recommendation performance with the RAND splitting strategy is consistently better than the performance with the SKEW splitting strategy on all three datasets. This shows that learning recommendation models when the training and testing distributions are different (SKEW) is a more challenging task than under the same train-test distribution (RAND).

The following observations are consistent no matter what the splitting strategy is. First, the recommendation models shown in Table \ref{tab:SKEW_results} and Table \ref{tab:RAND_results} can be categorized into classic association-based models and deconfounded models, where the difference between the two is whether the effects of confounders are mitigated or not. From the results, we can see that the deconfounded recommendation models (DecRS, IPW-MF, DCF, HCR and DCCF) achieve better recommendation performance than classic recommendation models (MF and DMF) in most cases on all three datasets. 
When averaging across all deconfounded recommendation models on all datasets using the SKEW splitting strategies, the deconfounded models achieve 54.3\% relative improvement on nDCG@5 than classic recommendation models, 55.0\% on Recall@5, and 51.7\% on Precision@5. 
When averaging across all deconfounded recommendation models on all datasets using the RAND splitting strategies, the deconfounded models achieve 4.9\% relative improvement on nDCG@5, 2.9\% on Recall@5, and 8.3\% on Precision@5. 
According to this comparison, we can observe that the existence of confounders will lead to inaccurate recommendations and deconfounded recommendation models will improve the recommendation performance by mitigating the effect of confounders. Meanwhile, comparing the improvement on both splitting strategies, the improvement brought by the deconfounded models is more significant on the SKEW datasets than the RAND datasets. 

\begin{table*}
\small
\begin{tabular}{l|ccc|ccc|ccc}
\toprule
    \multirowcell{2}{Methods} & \multicolumn{3}{c|}{Electronics} & \multicolumn{3}{c|}{CDs and Vinyl} & \multicolumn{3}{c}{Yelp} \\
     & nDCG@5 & Rec@5 & Pre@5 & nDCG@5 & Rec@5 & Pre@5 & nDCG@5 & Rec@5 & Pre@5 \\\midrule
    MF & 0.0201 & 0.0185 & 0.0170 & 0.1567 & 0.1158 & 0.1344 & 0.1013 & 0.0601 & 0.0980\\
    DMF &  0.0253 & 0.0237 & 0.0205 & 0.1363 & 0.1011 & 0.1164 & 0.1107 & 0.0666 & 0.1065\\
    DecRS & 0.0687 & 0.0652 & 0.0549 & 0.2046 & 0.1550 & 0.1697 & 0.1108 & 0.0676 & 0.1079\\
    IPW-MF & 0.0447 & 0.0398 & 0.0345 & 0.1959 & 0.1434 & 0.1622 & 0.1099 & 0.0666 & 0.1071\\
    DCF & 0.0317 & 0.0308 & 0.0257 & 0.1573 & 0.1162 & 0.1352 & 0.1115 & 0.0679 & 0.1083\\
    HCR & 0.0790 & 0.0751 & 0.0626 & 0.2033 & 0.1548 & 0.1707 & 0.1123 & 0.0680 & 0.1091\\
    % HKBE &  &  &  &  &  &  \\
    % FMG &  &  &  &  &  &  \\
    % Meta-Prod2vec &  &  &  &  &  &  \\ 
    \cdashline{1-10}
    DCCF$\_$NS & 0.0429 & 0.0413 & 0.0350 & 0.1581 & 0.1168 & 0.1330& 0.1087 & 0.0660 & 0.1058\\
    DCCF$\_$ND & 0.0413 & 0.0401 & 0.0329 & 0.1427 & 0.1078 & 0.1210 & 0.0991 & 0.0598 & 0.0961\\
    DCCF (ours) & \textbf{0.0934} & \textbf{0.0880} & \textbf{0.0737} & \textbf{0.2289} & \textbf{0.1702} & \textbf{0.1892} & \textbf{0.1184} & \textbf{0.0718} & \textbf{0.1143}\\
    \bottomrule
\end{tabular}
\caption{Recommendation performance for \textit{Electronics}, \textit{CDs and Vinyl}, and \textit{Yelp} using the SKEW splitting strategy. We evaluate on ranking task with metrics nDCG@5, Recall@5 and Precision@5. The best results are highlighted in bold. The improvements are significant at p < 0.01.}
\label{tab:SKEW_results}
% \vspace{-20pt}
\end{table*}

% \subsubsection{SKEW Splitting}

\begin{table*}
\small
\begin{tabular}{l|ccc|ccc|ccc}
\toprule
    \multirowcell{3}{Methods} & \multicolumn{3}{c|}{Electronics} & \multicolumn{3}{c|}{CDs and Vinyl} & \multicolumn{3}{c}{Yelp} \\
     & nDCG@5 & Rec@5 & Pre@5 & nDCG@5 & Rec@5 & Pre@5 & nDCG@5 & Rec@5 & Pre@5 \\\midrule
    MF & 0.1463 & 0.1454 & 0.1027 & 0.2891 & 0.2168 & 0.2382 & 0.2980 & 0.1776 & 0.2734\\
    DMF & 0.0948 & 0.0956 & 0.0681 & 0.1618 & 0.1208 & 0.1395 & 0.2483 & 0.1523 & 0.2355\\
    DecRS & 0.1589 & 0.1431 & 0.1200 & 0.3093 & 0.2361 & 0.2536 & 0.3055 & 0.1846 & 0.2819\\
    IPW-MF & 0.1502 & 0.1385 & 0.1135 & 0.2954 & 0.2197 & 0.2553 & 0.3034 & 0.1832 & 0.2802\\
    DCF & 0.1463 & 0.1340 & 0.1124 & 0.2885 & 0.2175 & 0.2389 & 0.2995 & 0.1817 & 0.2772\\
    HCR & 0.1613 & 0.1487 & 0.1224 & 0.3059 & 0.2320 & 0.2495 & 0.3045 & 0.1842 & 0.2804\\
    \cdashline{1-10}
    DCCF$\_$NS & 0.1542 & 0.1419 & 0.1171 & 0.2886 & 0.2164 & 0.2360 & 0.2999 & 0.1813 & 0.2766\\
    DCCF$\_$ND & 0.1387 & 0.1276 & 0.1045 & 0.2247 & 0.1692 & 0.1882 & 0.2887 & 0.1737 & 0.2665\\
    DCCF (ours) & \textbf{0.1742} & \textbf{0.1598} & \textbf{0.1312} & \textbf{0.3237 }& \textbf{0.2471} & \textbf{0.2643} & \textbf{0.3073} & \textbf{0.1861} & \textbf{0.2835}\\
    % random & 0.7168 & 0.4564 & 0.9472 & 0.7821 & 0.9255 & 0.6624 \\
    % max & 0.7017 & 0.4509 & 0.8309 & 0.6485 & 0.8541 & 0.5678 \\
    % max sample & 0.6527 & 0.4049 & 0.8345 & 0.6509 & 0.8490 & 0.5475\\\cdashline{1-7}
    % min & 0.7101 & 0.4860 & 0.9028 & 0.7422 & 0.8867 & 0.6554 \\
    % min sample & 0.7182 & 0.4913 & 0.9180 & 0.7573 & 0.8970 & 0.6500 \\\cdashline{1-7}
    % max diff & 0.7104 & 0.4433 & 0.8866 & 0.7024 & 0.8736 & 0.5974 \\
    % max diff sample & 0.7190 & 0.4536 & 0.9134 & 0.7432 & 0.8892 & 0.6118\\\cdashline{1-7}
    % min diff & 0.7379 & 0.4932 & 0.9554 & 0.8280 & 0.9048 & 0.6693 \\
    % min diff sample & 0.7527 & 0.4981 & 0.9459 & 0.7956 & 0.9079 & 0.6693 \\\cdashline{1-7}
    % No sample & 0.6781 & 0.4455 & 0.8488 & 0.6611 & 0.8847 & 0.6265\\
    % Ours &  &  & 0.8059 & 0.5708 & 0.9103 & 0.6461 \\
    % no Exp Loss &  &  & 0.7062 & 0.4828 &  & \\
    % Ablation &  &  & 0.6580 & 0.4045 &  & \\
    \bottomrule
\end{tabular}
\caption{Recommendation performance for \textit{Electronics}, \textit{CDs and Vinyl}, and \textit{Yelp} using the RAND splitting strategy. We evaluate on ranking task with metrics nDCG@5, Recall@5 and Precision@5. The best results are highlighted in bold. The improvements are significant at p < 0.01.}
\label{tab:RAND_results}
% \vspace{-20pt}
\end{table*}

Among deconfounded recommendation models, we can see that our DCCF model achieves the best recommendation performance in most cases. Compared with the strongest deconfounded model in baseline, when averaging across all datasets using the SKEW splitting strategies, our model yields 18.2\% improvement on nDCG@5, 16.9\% improvement on Recall@5, and 17.2\% improvement on Precision@5. When averaging across all datasets using the RAND splitting strategies, our model yields 4.9\% improvement on nDCG@5, 5.7\% improvement on Recall@5, and 4.4\% improvement on Precision@5.
These observations imply that by applying the front-door adjustment into the estimation of user's preference, our model is capable of reducing the effect of unobserved confounders and further improving the recommendation performance. Moreover, based on the results, applying the front-door adjustment is more effective than other deconfounded models.

Although the front-door adjustment is effective on reducing the effect of unobserved confounders, it requires inherent item features. We need to confirm that the improved performance is indeed brought by the deconfounding effect instead of the use of inherent item features.
% However, the improvement of our DCCF model is not brought by involving item feature representation into the model. 
Therefore, we consider a variant of our model DCCF$\_$ND, which also involves inherent item feature representation into estimation but without the front-door adjustment as deconfounding component. More specifically, unlike DCCF that applies the front-door adjustment with weighted sum over sampled items to obtain the estimation of a user-item pair, DCCF$\_$ND only considers the user-item pair itself when calculating the estimation. From the performance in Table \ref{tab:SKEW_results} and Table \ref{tab:RAND_results}, we can see that DCCF$\_$ND is even worse than classic models in many cases. Therefore, the improvement of our model is brought by deconfounding with the front-door adjustment instead of involving inherent item feature representation into calculation. We also include a variant of our model DCCF$\_$NS, which involves exposure probability but without sampled items. It can be considered as a reweighting method, instead of applying the front-door adjustment. From the performance in Table \ref{tab:SKEW_results} and Table \ref{tab:RAND_results}, we can see that DCCF$\_$NS is able to improve recommendation performance compared with classic models, and even achieve better performance than deconfounded models in some cases. However, it cannot obtain better performance than our model. Therefore, applying front-door adjustment is more effective than using exposure probability for reweighting.

\begin{table*}
\small
\begin{tabular}{l|ccc|ccc}
\toprule
    \multirowcell{2}{Methods} & \multicolumn{3}{c|}{\textbf{SD1}} & \multicolumn{3}{c}{\textbf{SD2}}\\
     & nDCG@5 & Rec@5 & Pre@5 & nDCG@5 & Rec@5 & Pre@5 \\\midrule
    MF & 0.1201 & 0.0833 & 0.1023 & 0.1069 & 0.0654 & 0.1167\\
    DMF & 0.1134 & 0.0837 & 0.0972 & 0.1090 & 0.0663 & 0.1195\\
    DecRS & 0.1219 & 0.0888 & 0.1030 & 0.1138 & 0.0973 & 0.1235\\
    IPW-MF & 0.1224 & 0.0899 & 0.1034 & 0.1176 & 0.0715 & 0.1276\\
    DCF & 0.1287 & 0.0913 & 0.1127 & 0.1254 & 0.0751 & 0.1361\\
    HCR & 0.1392 & 0.1000 & 0.1186 & 0.1334 & 0.0787 & 0.1424\\
    \cdashline{1-7}
    DCCF$\_$NS & 0.1212 & 0.0878 & 0.1066 & 0.0988 & 0.0602 & 0.1072\\
    DCCF$\_$ND & 0.1025 & 0.0748 & 0.0880 & 0.0844 & 0.0521 & 0.0923\\
    DCCF (ours) & \textbf{0.1438} & \textbf{0.1066} & \textbf{0.1216} & \textbf{0.1358} & \textbf{0.0807} & \textbf{0.1466}\\
    % random & 0.7168 & 0.4564 & 0.9472 & 0.7821 & 0.9255 & 0.6624 \\
    % max & 0.7017 & 0.4509 & 0.8309 & 0.6485 & 0.8541 & 0.5678 \\
    % max sample & 0.6527 & 0.4049 & 0.8345 & 0.6509 & 0.8490 & 0.5475\\\cdashline{1-7}
    % min & 0.7101 & 0.4860 & 0.9028 & 0.7422 & 0.8867 & 0.6554 \\
    % min sample & 0.7182 & 0.4913 & 0.9180 & 0.7573 & 0.8970 & 0.6500 \\\cdashline{1-7}
    % max diff & 0.7104 & 0.4433 & 0.8866 & 0.7024 & 0.8736 & 0.5974 \\
    % max diff sample & 0.7190 & 0.4536 & 0.9134 & 0.7432 & 0.8892 & 0.6118\\\cdashline{1-7}
    % min diff & 0.7379 & 0.4932 & 0.9554 & 0.8280 & 0.9048 & 0.6693 \\
    % min diff sample & 0.7527 & 0.4981 & 0.9459 & 0.7956 & 0.9079 & 0.6693 \\\cdashline{1-7}
    % No sample & 0.6781 & 0.4455 & 0.8488 & 0.6611 & 0.8847 & 0.6265\\
    % Ours &  &  & 0.8059 & 0.5708 & 0.9103 & 0.6461 \\
    % no Exp Loss &  &  & 0.7062 & 0.4828 &  & \\
    % Ablation &  &  & 0.6580 & 0.4045 &  & \\
    \bottomrule
\end{tabular}
\caption{Recommendation performance for two synthetic datasets \textbf{SD1} and \textbf{SD2}. We evaluate on ranking task with metrics nDCG@5, Recall@5 and Precision@5. The best results are highlighted in bold. The improvements are significant at p < 0.01.}
\label{tab:syn_results}
% \vspace{-20pt}
\end{table*}

For synthetic datasets, the recommendation performance for models are shown in Table \ref{tab:syn_results}. Dataset \textbf{SD1} simulates a scenario where the selected inherent item features completely intercept the causal effect from items to preference, while \textbf{SD2} represents a situation where the mediator intercepts a significant portion of the causal effect from items to preference. We can observe that deconfounded models still achieve better performance than classical models. Among different deconfounded models, DecRS and IPW-MF mainly focus on mitigating the effect of item popularity, therefore, they can only achieve lightly improved performance since the confounders in synthetic data are not necessary related to item popularity. The remaining deconfounded models (i.e., DCF, HCR and DCCF) are not restricted to a specific confounder, thus are capable of achieving better deconfounding performance. Our DCCF model achieves the best performance on both synthetic datasets, which illustrate the effectiveness of our model on mitigating the effect of unobserved confounders, even when the mediator cannot fully intercept the causal effect from items to preference.

\begin{table*}
\footnotesize
\begin{tabular}{l|ccc|ccc|ccc}
\toprule
    \multirowcell{2}{Methods} & \multicolumn{3}{c|}{Electronics} & \multicolumn{3}{c|}{CDs and Vinyl} & \multicolumn{3}{c}{Yelp}\\
     & nDCG@5 & Rec@5 & Pre@5 & nDCG@5 & Rec@5 & Pre@5 & nDCG@5 & Rec@5 & Pre@5\\\midrule
    DCCF$\_$Random & 0.0340 & 0.0333 & 0.0272 & 0.0831 & 0.0613 & 0.0714 & 0.0722 & 0.0418 & 0.0695\\
    DCCF$\_$Uniform & 0.0338 & 0.0334 & 0.0272 & 0.0853 & 0.0634 & 0.0723 & 0.0717 & 0.0411 & 0.0691\\
    DCCF$\_$Bias & 0.0687 & 0.0652 & 0.0549 & 0.2053 & 0.1535 & 0.1700 & 0.1106 & 0.0666 & 0.1071\\
    DCCF$\_$Unbias & \textbf{0.0846} & \textbf{0.0814} & \textbf{0.0670} & \textbf{0.2167} & \textbf{0.1654} & \textbf{0.1806} & \textbf{0.1124} & \textbf{0.0682} & \textbf{0.1090}\\
    % no Exp Loss &  &  & 0.7062 & 0.4828 &  & \\
    % Ablation &  &  & 0.6580 & 0.4045 &  & \\
    \bottomrule
\end{tabular}
\caption{The recommendation performance of DCCF under different exposure models on three datasets using the SKEW splitting strategy. We report the evaluatation on ranking task with metrics nDCG@5, Recall@5 and Precision@5. The best results are highlighted in bold.}
\label{tab:SKEW_albation}
\vspace{-10pt}
\end{table*}

\begin{table*}
\footnotesize
\begin{tabular}{l|ccc|ccc|ccc}
\toprule
    \multirowcell{2}{Methods} & \multicolumn{3}{c|}{Electronics} & \multicolumn{3}{c|}{CDs and Vinyl} & \multicolumn{3}{c}{Yelp}\\
     & nDCG@5 & Rec@5 & Pre@5 & nDCG@5 & Rec@5 & Pre@5 & nDCG@5 & Rec@5 & Pre@5 \\\midrule
    DCCF$\_$Random & 0.0609 & 0.0581 & 0.0469 & 0.1194 & 0.0867 & 0.1008 & 0.1332 & 0.0799 & 0.1277\\
    DCCF$\_$Uniform & 0.0627 & 0.0598 & 0.0484 & 0.1199 & 0.0862 & 0.1005 & 0.1428 & 0.0860 & 0.1362\\
    DCCF$\_$Bias & 0.1510 & 0.1362 & 0.1139 & 0.2886 & 0.2164 & 0.2360 & 0.2876 & 0.1743 & 0.2673\\
    DCCF$\_$Unbias & \textbf{0.1583} & \textbf{0.1422} & \textbf{0.1194} & \textbf{0.2939} & \textbf{0.2290} & \textbf{0.2428} & \textbf{0.2958} & \textbf{0.1786} & \textbf{0.2733}\\
    % no Exp Loss &  &  & 0.7062 & 0.4828 &  & \\
    % Ablation &  &  & 0.6580 & 0.4045 &  & \\
    \bottomrule
\end{tabular}
\caption{The recommendation performance of DCCF under different exposure models on three datasets using the RAND splitting strategy. We report the evaluatation on ranking task with metrics nDCG@5, Recall@5 and Precision@5. The best results are highlighted in bold.}
\label{tab:RAND_albation}
% \vspace{-20pt}
\end{table*}

\begin{figure*}[t!]
\captionsetup[sub]{font=small,labelfont=normalfont,textfont=normalfont}
    \centering
    \begin{subfigure}{0.45\textwidth}
        \includegraphics[scale=0.45]{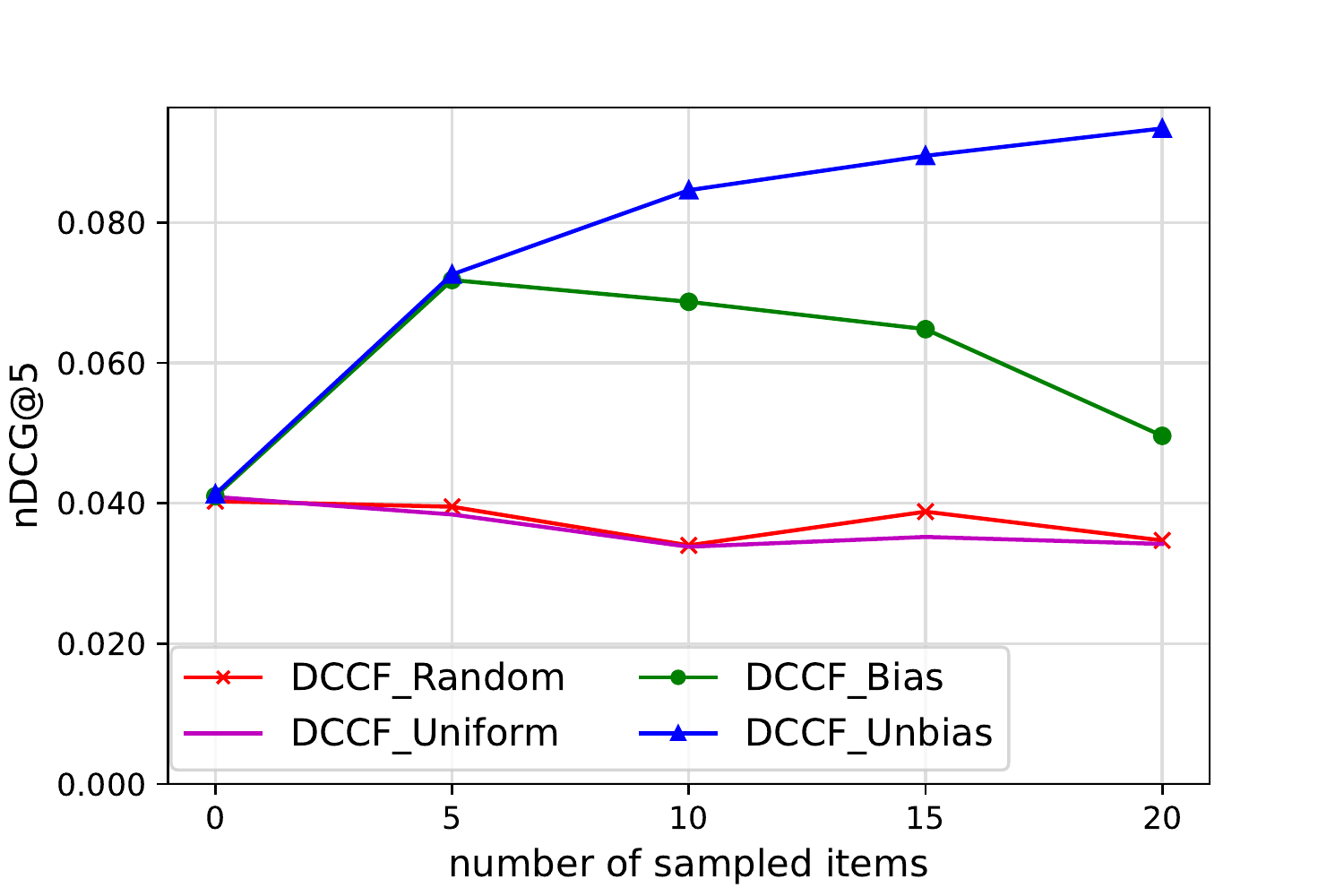}
        \caption{Electronics (SKEW)}
    \end{subfigure}
    \begin{subfigure}{0.45\textwidth}
        \includegraphics[scale=0.45]{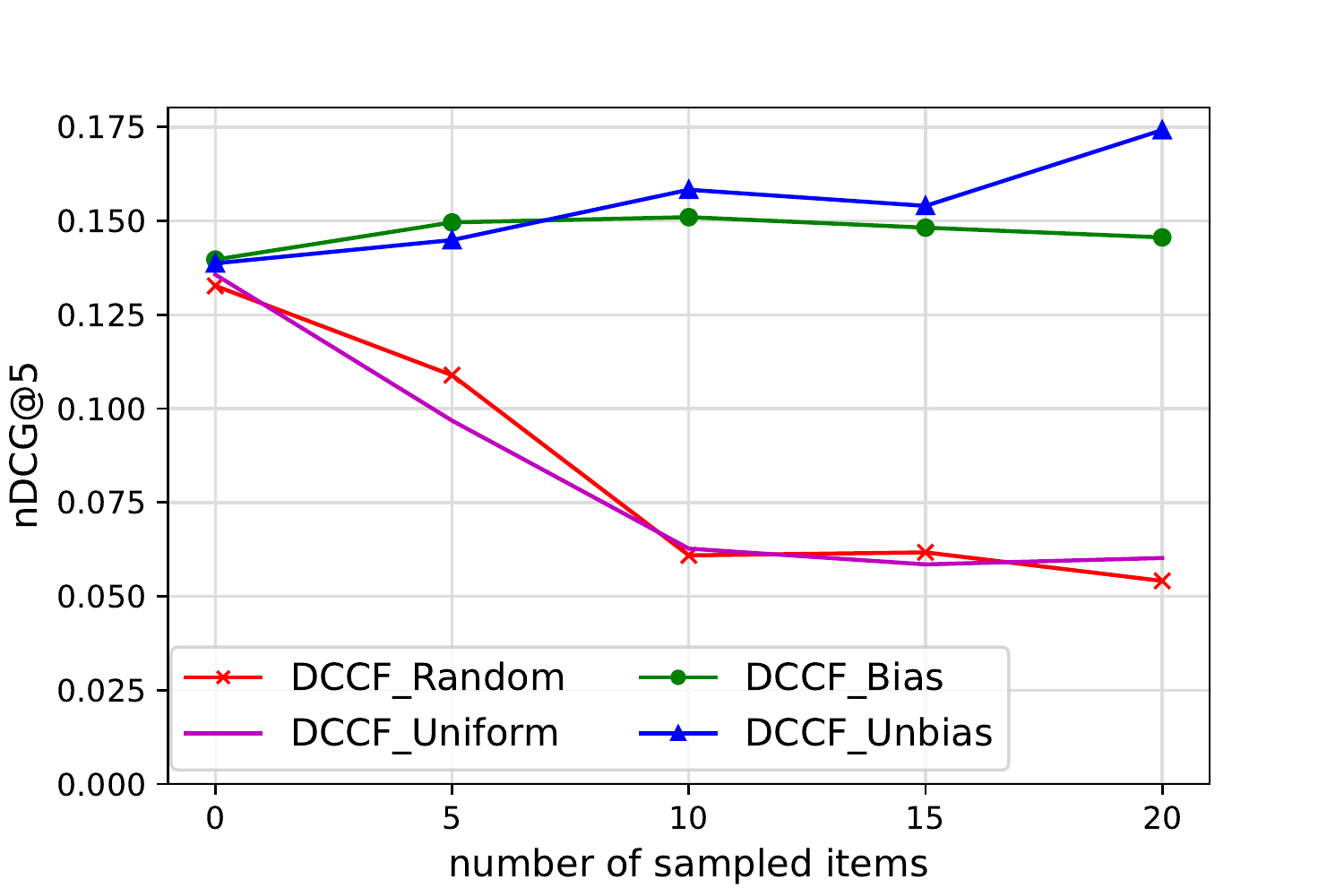}
        \caption{Electronics (RAND)}
    \end{subfigure}
    \begin{subfigure}{0.45\textwidth}
        \includegraphics[scale=0.45]{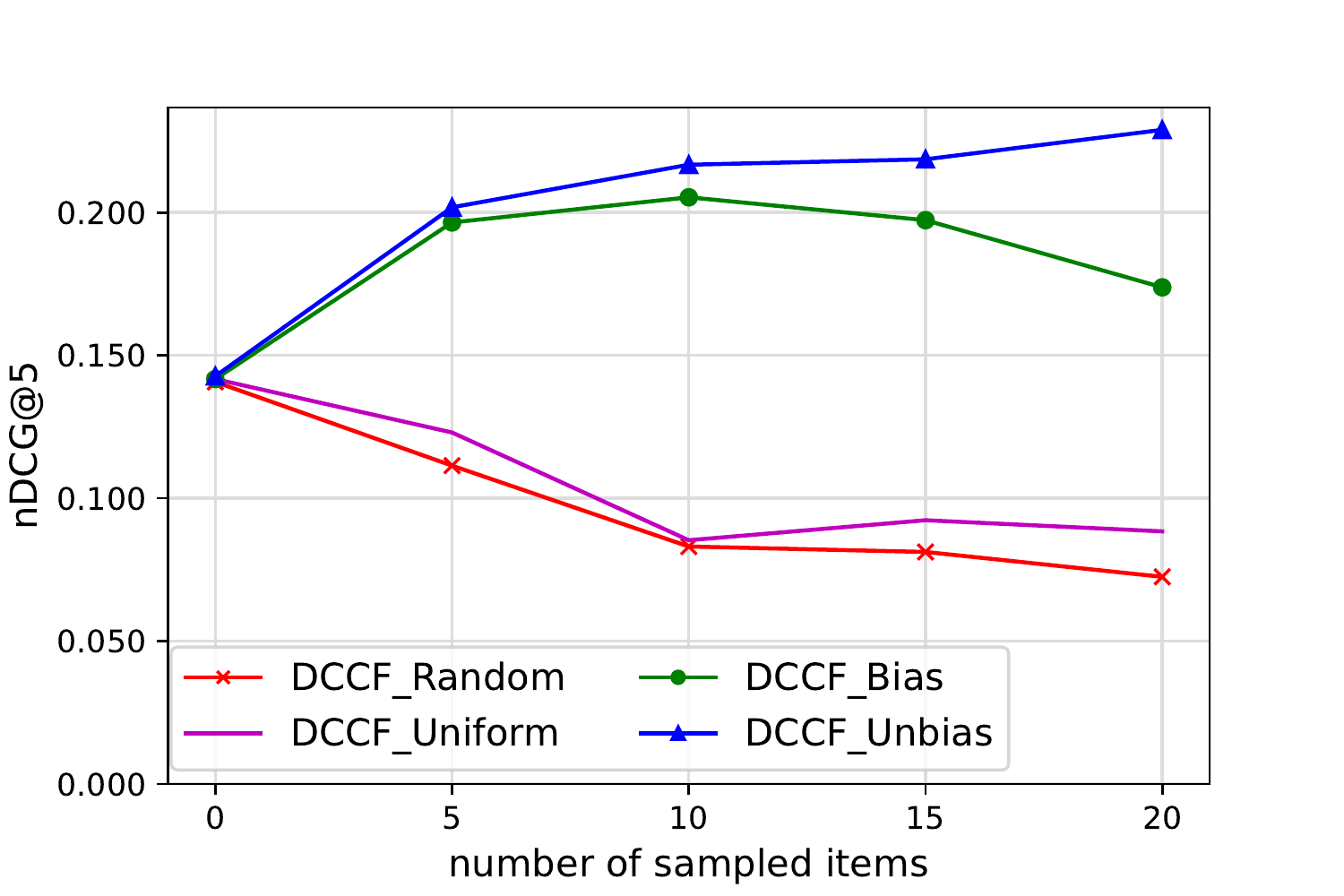}
        \caption{CDs and Vinyl (SKEW)}
    \end{subfigure}
    \begin{subfigure}{0.45\textwidth}
        \includegraphics[scale=0.45]{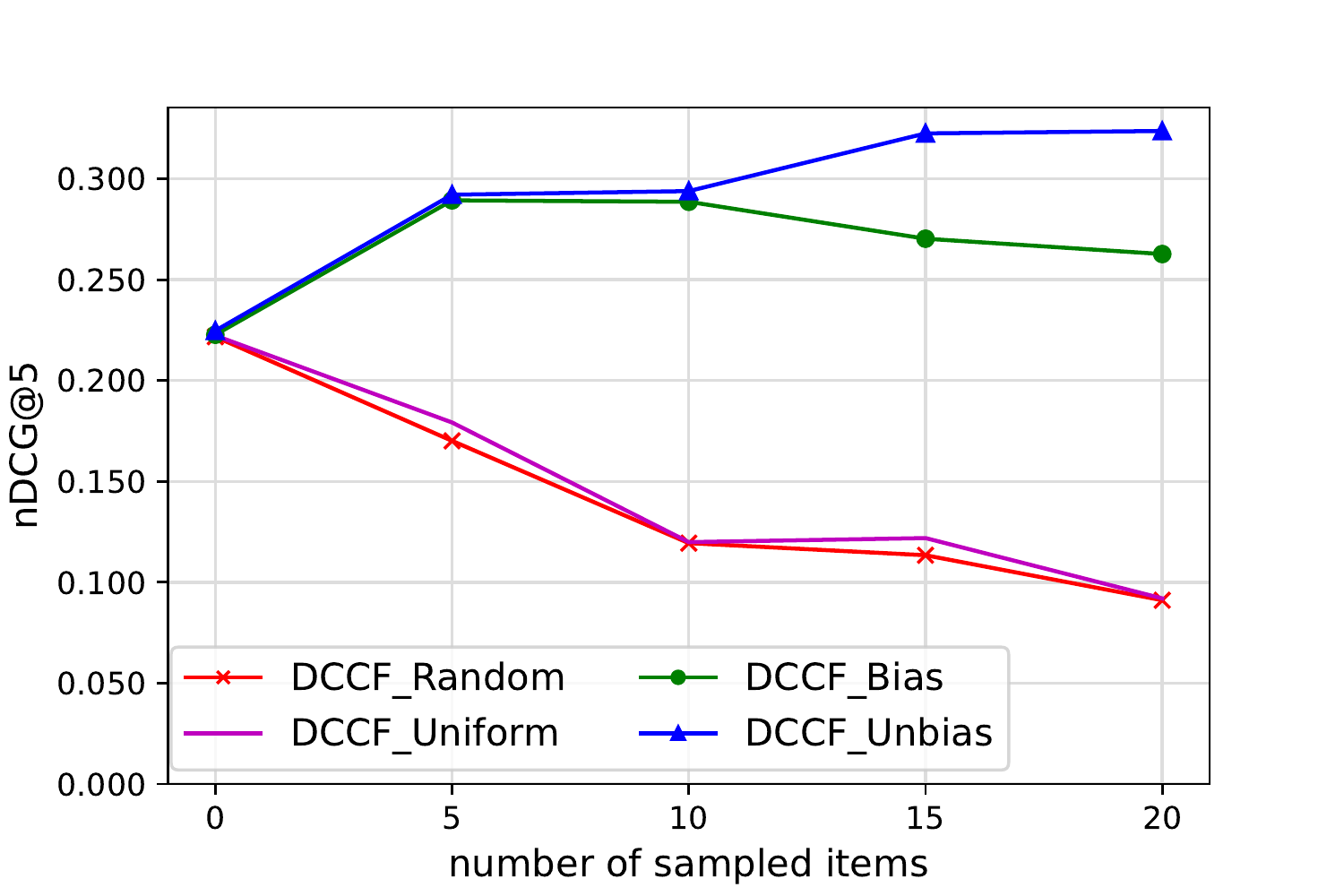}
        \caption{CDs and Vinyl (RAND)}
    \end{subfigure}
    \begin{subfigure}{0.45\textwidth}
        \includegraphics[scale=0.45]{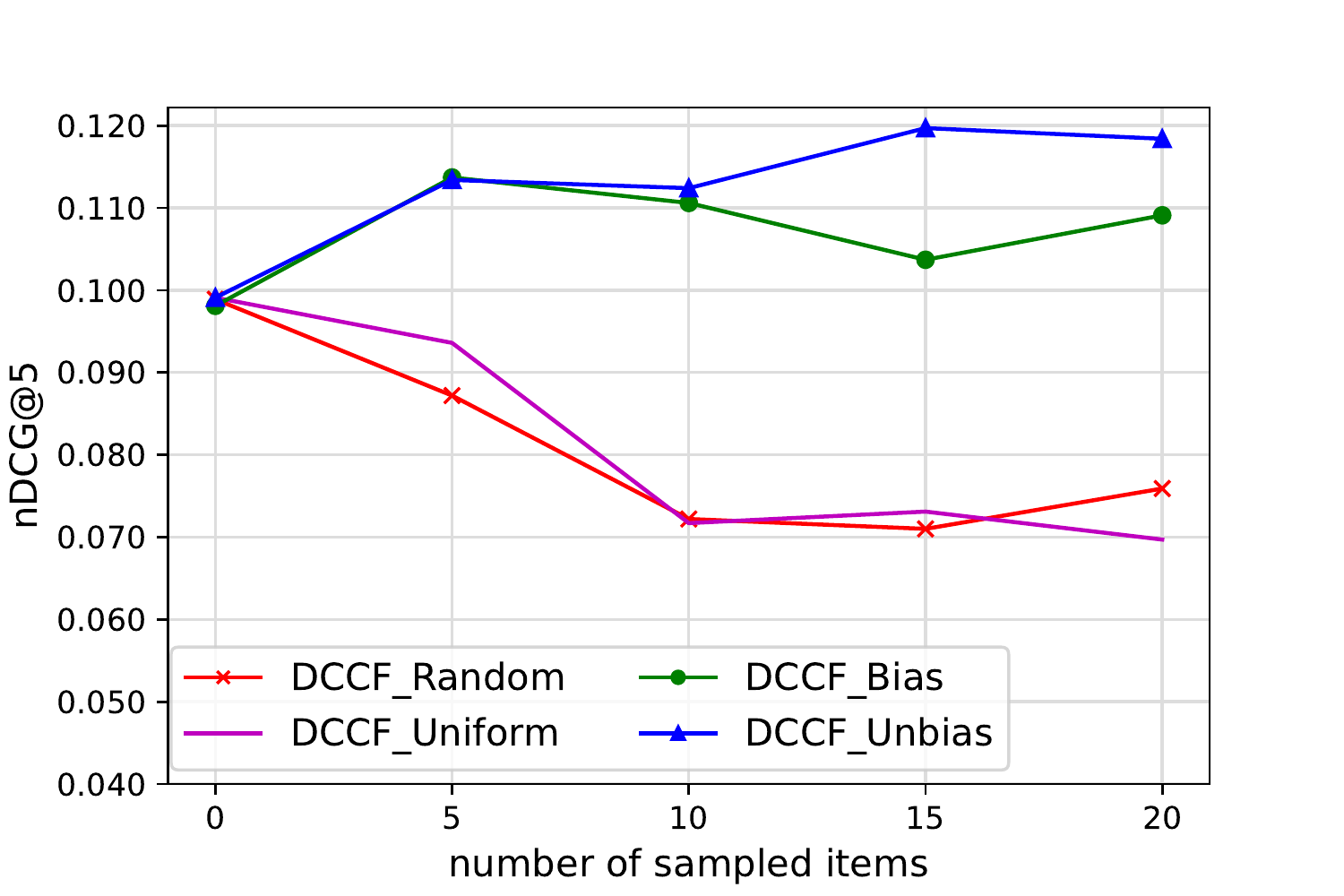}
        \caption{Yelp (SKEW)}
    \end{subfigure}
    \begin{subfigure}{0.45\textwidth}
        \includegraphics[scale=0.45]{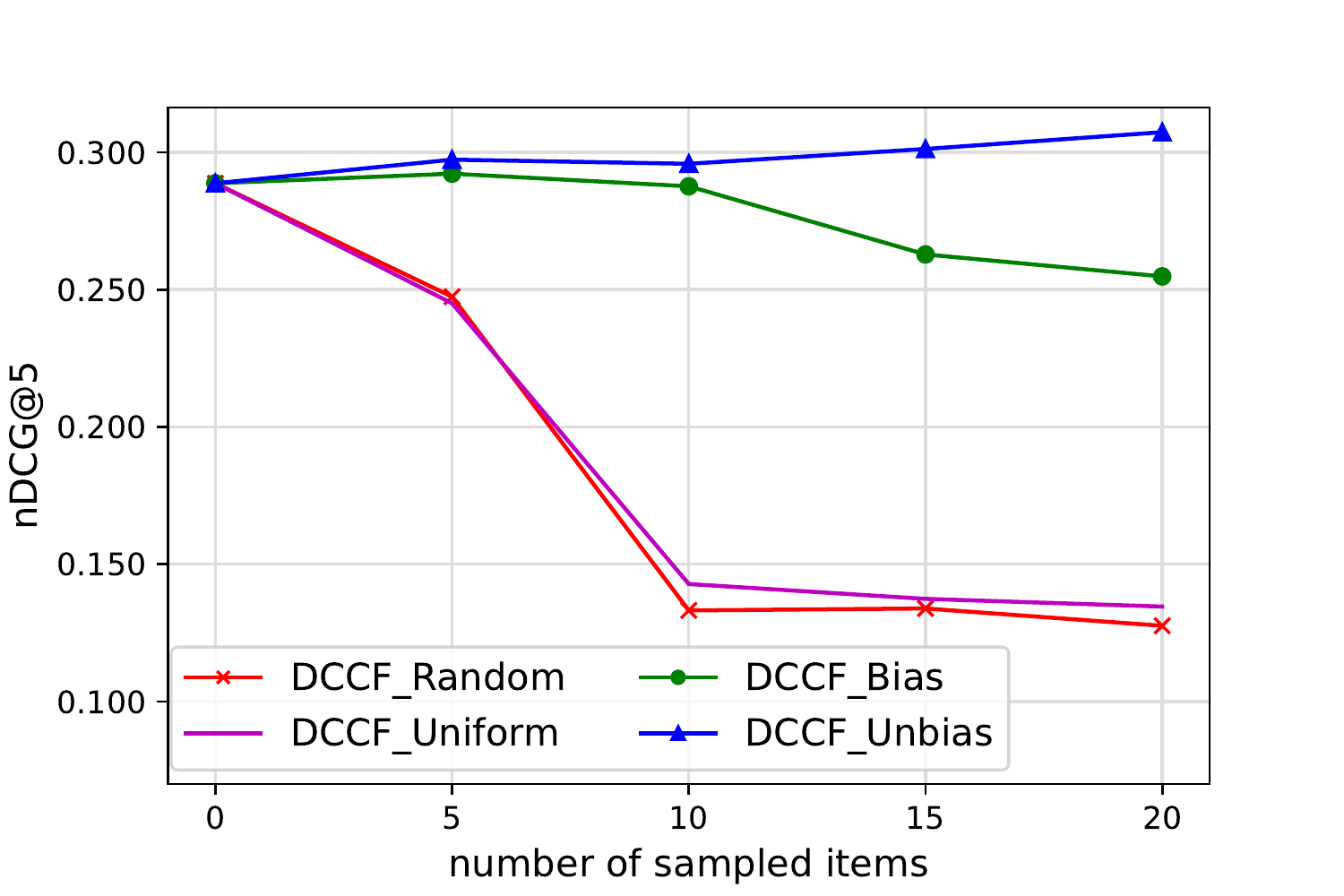}
        \caption{Yelp (RAND)}
    \end{subfigure}
    % \vspace{-10pt}
    \caption{nDCG@5 for different exposure probability models under different numbers of sampled items.}
    \label{fig:expo}
    % \vspace{-5pt}
\end{figure*}

\subsection{Influence of the Number of Sampled Items}
In this section, we will discuss the influence of the number of sampled items to answer \textbf{RQ3}. As we mentioned in Section \ref{sec:preference}, instead of strictly following the front-door adjustment as Eq.\eqref{eq:frontdoormcs}, we randomly sample a set of items and estimate the preference by the sample-based front-door adjustment (i.e., as Eq.\eqref{eq:samplefrontdoor}). 

To discuss how the number of sampled items influence the recommendation performance, in Figure \ref{fig:expo}, we plot nDCG@5 for DCCF (i.e., DCCF$\_$Unbias in Figure \ref{fig:expo}) with different numbers of sampled items. We can observe that the recommendation performance increases with the increasing of the number of sampled items, but the magnitude of the increment decreases as the number of sampled items increases. Considering the model applies a two layer MLP with dimension $D_{uv}$, the time complexity of calculating Eq.\eqref{eq:conditionalpref} would be $(D_{uv}+D_m)D_{uv} + 2D_{uv}^2$. Suppose $D_m$ is linear to $D_{uv}$, the time complexity is $O(D_{uv}^2)$. For a fixed number of sampled inherent item feature $d$ ($d\ll D_{uv}$), if the number of sampled item $n$ is small enough (i.e., $n\ll D_{uv}$), the time complexity of calculating $P(y|u,do(v))$ is $O(D_{uv}^2)$. However, if $n$ is large (i.e., linear to $D_{uv}$), the time complexity would be $O(D_{uv}^3)$. Additionally, larger $n$  will occupy more memory during the training. Therefore, it is important to choose an appropriate value of $n$ to balance the recommendation performance and efficiency when applying the DCCF model.

\subsection{Influence of Exposure Models}\label{sec:expomodels}
In this section, we will discuss the influence of the exposure models to answer \textbf{RQ4}. As we mentioned before, the exposure probability, which is required by the front-door adjustment but not available in the feedback data, is an essential component of our model. To empirically show the influence of it, we design three versions of our model with different exposure probabilities. The only difference among them is how to obtain the exposure probability.
\begin{itemize}[leftmargin=*]
    \item \textbf{DCCF$\_$Random}: In this model, the exposure probabilities are not learned from the data, instead, we randomly generate a matrix to represent the exposure probability.
    \item \textbf{DCCF$\_$Uniform}: In this model, we assume that each item has an equal probability to be exposed for each user.
    \item \textbf{DCCF$\_$Bias}: In this model, we estimate the exposure probability as Eq.\eqref{eq:biasedMF}, and use the pair-wise learning method to train the model based on implicit feedback.
    \item \textbf{DCCF$\_$Unbias}: This model is the same as \textbf{DCCF} in Table \ref{tab:SKEW_results} and Table \ref{tab:RAND_results}. Concretely, we consider the bias in the implicit feedback data, and apply Eq.\eqref{eq:ips_expo} to obtain unbiased estimation of exposure probabilities.
\end{itemize}

Following the setting in main experiment, we evaluate three datasets using two splitting strategies on ranking task with metrics nDCG@5, Recall@5 and Precision@5. The recommendation performance on the SKEW splitting strategy is shown in Table \ref{tab:SKEW_albation} and the performance on the RAND splitting strategy is shown in Table \ref{tab:RAND_albation}. For each of the version, we use the same representation of inherent item features and sample 10 items for each pair to calculate the estimated preference. 

From the results, we can see that DCCF$\_$Unbias achieves the best performance; DCCF$\_$Random and DCCF$\_$Uniform have the worst performance (DCCF$\_$Random and DCCF$\_$Uniform get similar performance) in most cases. This observation is consistent on both RAND and SKEW splitting strategies. Comparing the four versions with different exposure models shows that accurate exposure probability will lead to accurate estimation under the front-door adjustment.
% To some extent, the comparison among the three versions with different exposure models can support this claim. 
Specifically, the exposure probability is not observed in the feedback data, therefore, we need to learn a model to estimate it from observational feedback data. DCCF$\_$Unbias uses the unbiased estimation and gets the most accurate exposure probability, while DCCF$\_$Random randomly generates exposure probabilities and DCCF$\_$Uniform applyies uniform probabilities, which are irrelevant with the feedback data, and thus will get the least accurate probability. We can see that the ranking of the  exposure probability accuracy matches the ranking of recommendation performance.

The exposure model affects our model in several ways, i.e., not only the best recommendation performance, but also the trend of the recommendation performance as the number of sampled item increases. As we mentioned in Section \ref{sec:model} and Eq.\eqref{eq:finalestimation}, we apply the front-door adjustment over a set of sampled items to make it suitable for the recommendation scenario. Ideally, if there exist the ground truth values of exposure probability and we apply them into the calculation, then increasing the number of sampled items may reduce the efficiency but will at least not hurt the performance significantly. 

In order to investigate how our model changes with increasing the number of sampled items under different exposure models, we plot nDCG@5 for four versions of the model with different numbers of sampled items,
% from 0 to 20 
as shown in Figure \ref{fig:expo}. 
% When the number of sampled items is set to 0, all versions of the model degenerate to DCCF$\_$ND (i.e., as in Table \ref{tab:SKEW_results} and Table \ref{tab:RAND_results}).
% Considering the efficiency of the experiment, we sample up to 20 items. 
As we can see from the figure, the performance of DCCF$\_$Unbias increases with the increasing of sample number, the performance of DCCF$\_$Bias increases first and then drops after 5 samples, and the performance of DCCF$\_$Random and DCCF$\_$Uniform is monotonically decreasing. Meanwhile, for a given number of sampled items, the performance among the four models follows the conclusion as Table \ref{tab:SKEW_albation} and Table \ref{tab:RAND_albation} in most cases, i.e., DCCF$\_$Unbias $>$ DCCF$\_$Bias $>$ DCCF$\_$Random/DCCF$\_$Uniform. 

For DCCF$\_$Unbias, since it has the most accurate exposure probability, it will achieve better performance with more sampled items. For DCCF$\_$Bias, the probability is also learned from data as DCCF$\_$Unbias but not accurate enough due to ignoring the bias. Therefore, when the number of sampled items is small, the model can get slightly better performance with the help of sampled items. However, when more sampled items are involved into the estimation, inaccurate probabilities may mislead front-door adjustment and hurt the performance, thus resulting in performance drops. For DCCF$\_$Random/DCCF$\_$Uniform, the sampled items do not provide useful information but instead introduce noises that completely mislead the front-door adjustment, and thus hurt the performance. In summary, accurate exposure probability will improve the performance while inaccurate probability hurts the performance. Therefore, it is important to adopt an accurate exposure model to obtain better performance.

% \vspace{-1ex}
\section{Conclusions and Future Work}\label{sec:conclusion}
In this paper, we notice that the unobserved confounders may result in inaccurate recommendations. To solve this problem, we propose a deconfounded causal collaborative filtering (DCCF) model based on front-door adjustment. Specifically, we first design a causal graph to depict user behaviors with unobserved confounders. We then apply the front-door adjustment to design the model for mitigating the influence of unobserved confounders. Considering the large scale of items in real-world system, we use the sample-based front-door adjustment to calculate the deconfounded estimation of users' preference. Experiments on real-world datasets with both skewed splitting and random splitting show that our deconfounded model can outperform both classical association-based models and deconfounded recommendation models. Furthermore, the experiments on the influence of exposure models show that the performance of our DCCF model is affected by the accuracy of the learned exposure probabilities. Therefore, a proper exposure model helps improve the recommendation performance of our deconfounded recommendation model.

% In this paper, we proposed FDR, a general framework to generate deconfounded recommendation. Unlike the most previous recommendation models that focus on extracting overall associative relationships between the users and the items, which suffer from potential bias. FDR tries to solve the recommendation problem from a causal perspective by predicting the user's satisfaction if a certain item is actually recommended. Since the confounders might not always be known and accountable, FDR applies front-door adjustment on the well-designed causal graph to filter out the both known and unknown bias, which leads to not only a better recommendation performance than the traditional associative-based recommender systems, but also outperforms the existing debiased recommendation baselines which only consider some certain known bias in the systems. We believe this is a good attempt to explore causal relationships in the recommendation field.

\textbf{Limitations and future work}. Our model treats all confounders as bias terms and removes all of their influence from the recommender system. However, the confounders may not always be harmful from the provider side. In practice, certain confounders can be useful and their effects should be remained in the recommendation. 
This requires a mixed recommendation strategy that considers partially associative rules and partially causal rules. 
The causal model used in DCCF could also be adapted to increase the interpretability of the recommender systems, as another important aspect to be considered for recommendation.


\section*{Acknowledgment}
This work was supported in part by NSF IIS-1910154,  IIS-2007907, IIS-2046457 and IIS-2127918. Any opinions, findings, conclusions or recommendations expressed in this material are those of the authors and do not necessarily reflect those of the sponsors.

\bibliographystyle{ACM-Reference-Format}
\bibliography{reference}

\end{document}